\pdfoutput=1

\documentclass[12pt, twoside, english]{article}
\usepackage[utf8]{inputenc}            
\usepackage[T1]{fontenc} 

\usepackage{mathtools}			
\usepackage{amssymb}
\usepackage{a4wide}
\usepackage{graphicx}
\usepackage{fancyhdr}
\usepackage{amsmath, amssymb}
\usepackage{subfigure}
\usepackage{setspace}
\usepackage{verbatim} 
\usepackage[english]{babel}
\usepackage[novbox]{pdfsync} 
\usepackage{caption}
\usepackage{float}
\usepackage{booktabs}

\usepackage{xfrac}

\usepackage{units}

\makeatletter
\usepackage{hyperref}

\usepackage{xcolor}

\hypersetup{
	colorlinks = false,
	linktocpage = true,
	allbordercolors = {0.8 0.8 0.8}
}

\addto\captionsenglish{}
\addto\captionsenglish{}
\addto\extrasenglish{}
\addto\extrasenglish{}
\addto\extrasenglish{}
\addto\extrasenglish{}
\addto\extrasenglish{}
\addto\extrasenglish{}
\addto\extrasenglish{}

\newcommand{\lyxaddress}[1]{
\par {\raggedright #1
\vspace{1.4em}
\noindent\par}
}

\numberwithin{equation}{section}

\newcommand{\vek}[1]{\mathchoice{\displaystyle\boldsymbol#1}
{\textstyle\boldsymbol#1}{\scriptstyle\boldsymbol#1}
{\scriptscriptstyle\boldsymbol#1}}
\newcommand{\mat}[1]{\mathchoice{\displaystyle\mathbf#1}
{\textstyle\mathbf#1}{\scriptstyle\mathbf#1}
{\scriptscriptstyle\mathbf#1}}

\renewcommand{\d}{ \ensuremath{\mathrm{d} }}

\makeatother

\begin{document}

\title{Simultaneous Analysis of Continuously Embedded Reissner--Mindlin Shells in 3D Bulk Domains}

\author{Michael Wolfgang Kaiser, Thomas-Peter Fries}
\maketitle

\lyxaddress{\begin{center}
Institute of Structural Analysis\\
Graz University of Technology\\
Lessingstr. 25/II, 8010 Graz, Austria\\
\texttt{www.ifb.tugraz.at}\\
\texttt{michael.kaiser@tugraz.at}
\end{center}}

\begin{abstract}
A mechanical model and numerical method for the \emph{simultaneous} analysis of Reissner-Mindlin shells with geometries implied by a continuous set of level sets (isosurfaces) over some three-dimensional bulk domain is presented. A three-dimensional mesh in the bulk domain is used in a tailored FEM formulation where the elements are by no means conforming to the level sets representing the shape of the individual shells. However, the shell geometries are bounded by the intersection curves of the level sets with the boundary of the bulk domain so that the boundaries are meshed conformingly. This results in a method which was coined Bulk Trace FEM before. The simultaneously considered, continuously embedded shells may be useful in the structural design process or for the continuous reinforcement of bulk domains. Numerical results confirm higher-order convergence rates.
\end{abstract}
\newpage{}\tableofcontents{}\newpage{}

\section{Introduction\label{sec:Introduction}}

Curved structures, e.g., shells and membranes, occur in a great variety in nature and engineering. Herein, the focus is on shells that are used in a wide range of engineering disciplines, e.g., civil, mechanical, aerospace and biomedical engineering \cite{Calladine_1983a, Zingoni_2018a, Basar_1985a}. There is an ongoing strong interest in mechanical models for shells, including their mathematical solution and approximation. Finite element analysis plays an important role in the simulation of the physical behaviour of structural shells, e.g., \cite{Bischoff_2017a, Chapelle_2011a, Zienkiewicz_2014a}.\\
\\
A major distinction for shell models which are based on the middle surface of the shell, relates to the consideration of transverse shear deformations. The \emph{Kirchhoff--Love} shell model neglects such shear deformations and is applicable for \emph{thin} shells. Its governing equation may be formulated by a fourth-order partial differential equation (PDE) with the displacements of the shell's middle surface as unknowns \cite{Basar_1985a, Bischoff_2017a, Love_1888a}. Numerical methods for this shell model require $C_1$-continuity due to the variational index 2 \cite{Echter_2013a}. Therefore, isogeometric analysis is often applied for the numerical analysis, being a modern variant of the FEM featuring improved continuity, e.g., \cite{Schoellhammer_2019a, Kiendl_2009a}. On the other hand, the \emph{Reissner--Mindlin} shell theory considers shear deformations \cite{Reissner_1945a, Basar_1985a} and is valid for thin and moderately thick shells \cite{Schoellhammer_2019b}. The governing equations are typically posed by two PDEs for the unknown middle surface displacements and rotations featuring the variational index is 1 \cite{Echter_2013a}. Consequently, classical FEMs, e.g., those based on Lagrange-type elements resulting in $C_0$-continuous function spaces, may be used in the analysis. A drawback of the Reissner--Mindlin shell model in the context of an FEM analysis is the sensitivity for locking effects, in particular when applied to rather thin shells \cite{Bischoff_2017a, Zou_2020a}. Another approach is to model shells as complete three dimensional bodies without  specific reference to the shell's middle surface \cite{Bischoff_2017a}.\\
\\
For those models using the shell's middle surface, the curved, two-dimensional surface, being the geometry of interest, is a manifold of co-dimension 1 embedded in the three-dimensional space $\mathbb{R}^3$ \cite{Schoellhammer_2019a}. This manifold can be described explicitly, e.g., by a parametrization, or implicitly, e.g., by the level-set method. The first variant of an \emph{explicit} geometry definition is the standard case in classical shell mechanics, cf. \cite{Basar_1985a, Bischoff_2017a}. There, the curved shell geometry in $\mathbb{R}^3$ results from a map of some two-dimensional parameter space $\boldsymbol{\hat{\Omega}} \in \mathbb{R}^2$. Through this parametrization, a (local) curvilinear coordinate system is introduced and geometric quantities, e.g., normal vectors, tangential vectors, and differential operators acting on the surface, are defined. These geometric quantities and differential operators are important ingredients in the governing PDEs characterizing the shell's physical behaviour \cite{Calladine_1983a, Basar_1985a}. Most finite element formulations for shells are based on curvilinear coordinates: For the Kirchhoff--Love shell model \cite{Kiendl_2009a, Nguyen-Thanh_2015a}, the Reissner--Mindlin shell model \cite{Chapelle_2011a, Dornisch_2013a, Dornisch_2014a, Benson_2010a}, the 7-parameter shell model \cite{Echter_2013a}, and \cite{Bischoff_2017a} for an overview. In an \emph{implicit} geometry description, the middle surface can be described by the level-set method  \cite{Osher_2006a,Fries_2017a,Fries_2017b}. It is then often given by the \emph{zero}-isosurface of a scalar level-set function $\phi$ \cite{Delfour_1995a, Schoellhammer_2019b}. Because in this case no curvilinear coordinates on the level set are present, a coordinate-free formulation of the governing equations of the shell model is required. This can be achieved by defining geometric quantities and differential operators with respect to a global Cartesian coordinate system using the Tangential Differential Calculus (TDC) \cite{Delfour_1996a, Delfour_2011a}. The resulting formulations are then applicable to explicit and implicit geometry definitions \cite{Fries_2020a, Schoellhammer_2021a}. Kirchhoff--Love shells have been reformulated based on the TDC in \cite{Schoellhammer_2019a, Delfour_1995a}, Reissner--Mindlin shells in \cite{Schoellhammer_2019b}, and a unified approach for ropes and membranes in \cite{Fries_2020a}. Furthermore, linear membranes are considered in \cite{Hansbo_2014a}, one-dimensional beams in a flat, two-dimensional plane in \cite{Kaiser_2023a} and in the three-dimensional space in \cite{Hansbo_2014b}. For a general overview on finite element methods for PDEs on surfaces see, e.g., \cite{Dziuk_2013a}.\\
\\
In a typical shell analysis, \emph{one} particular shell geometry is of interest. For several geometries, e.g., in a structural design context, \emph{successive} analyses on (slightly) modified geometries are common. Herein, in contrast, we propose to \emph{simultaneously} analyse a set of shell geometries embedded in some bulk domain. This can be useful for some types of application as, e.g., optimization problems or continuously embedded shells reinforcing some three-dimensional bulk material. The geometries of the shells are defined by \emph{a set of level sets} implied by a level-set function (instead, of only \emph{one} concrete level set such as the zero-isosurface). It turns out that the mechanical fields such as the displacements vary smoothly between the neighbouring level sets. It is then possible to formulate a mechanical model for \emph{all} shells embedded in a (prescribed) bulk domain and to analyse these structures \emph{simultaneously}. The aim of this paper is to formulate such a mechanical model and to show how its solution can be approximated using the FEM. The linear Reissner--Mindlin shell model, including shear deformations, is considered here. For the numerical analysis, the three-dimensional bulk domain is discretized by higher-order, three-dimensional Lagrange-type elements. These elements are conforming to the boundary of $\Omega$, however, they are not aligned to the level sets which describe the shape of the considered shells. Therefore, one may see this approach as a \emph{hybrid} of the classical Surface FEM and fictitious domain methods, e.g., the Trace FEM. Hence, it may be labelled as \emph{Bulk Trace FEM} as done in \cite{Fries_2023a}.\\
\\
Other formulations where models are defined on all level sets have been used in transport problems and  diffusion on stationary surfaces \cite{Dziuk_2008a} and on evolving surfaces \cite{Dziuk_2010a}. In \cite{Burger_2009a}, the simultaneous solution of some elliptic PDEs on all level sets within a bulk domain is compared to phase-field methods. A narrow-band method results when the bulk domain is minimized around a small number of selected level sets of interest \cite{Deckelnick_2010a,Deckelnick_2014a}. Narrow-band methods using finite differences for transport problems are described in \cite{Bertalmio_2001a, Greer_2006a, Greer_2006b}. Fictitious domain methods (Trace FEM, Cut FEM) may be used when only a \emph{single} level set is of interest, cf. \cite{Olshanskii_2014a, Olshanskii_2017a, Chernyshenko_2015a, Grande_2016a, Burman_2015a, Burman_2018a, Cenanovic_2016a}. In \cite{Cenanovic_2016a} Cut FEM is applied to linear membranes, in \cite{Schoellhammer_2021a}
Trace FEM is used for Reissner--Mindlin shells, for Kirchhoff-Love
shells a fictitious domain method is proposed in \cite{Gfrerer_2018a}, and geometrically non-linear ropes and membranes are considered in \cite{Fries_2020a}. In \cite{Fries_2023a}, the authors of this paper introduced an analogous approach to that proposed herein for geometrically non-linear ropes and membranes. Hence, this work may be seen as the first extension of \cite{Fries_2023a} to models in structural mechanics where out-of-plane bending is included.\\
\\
The paper is structured as follows: Section \ref{sec:LSinBD} introduces the geometrical foundations of the Bulk Trace FEM and the employed differential operators. In Section \ref{sec:RMSmodel}, the mechanical model, which describes the Reissner--Mindlin shell, is shown and the according weak form for the Bulk Trace FEM is derived in Section \ref{sec:RMS-BTF}. The discrete weak form including the enforcement of essential boundary conditions and the tangentiality constraint of the difference vector, describing the rotation of the shell, are discussed in Section \ref{sec:Impl}. Numerical results are given in Section \ref{sec:NumRes}. Some test cases are compared to classical benchmark tests for single shells, others confirm higher-order convergence rates based on residual errors and the stored energy error. Section \ref{sec:Concl} summarizes the paper with a conclusion and outlook.

\section{Level sets in a bulk domain}\label{sec:LSinBD}

The geometrical setup of two-dimensional level sets embedded in a three-dimensional bulk domain is introduced in this section. Furthermore, differential operators are defined which are later used in the formulation of the mechanical model and its numerical approximation.

\subsection{Geometric setup}\label{subsec:GeomSetup}

\begin{figure}
	\centering
	
	\subfigure[$\Omega$ and $\phi$]{\includegraphics[width=0.24\textwidth]{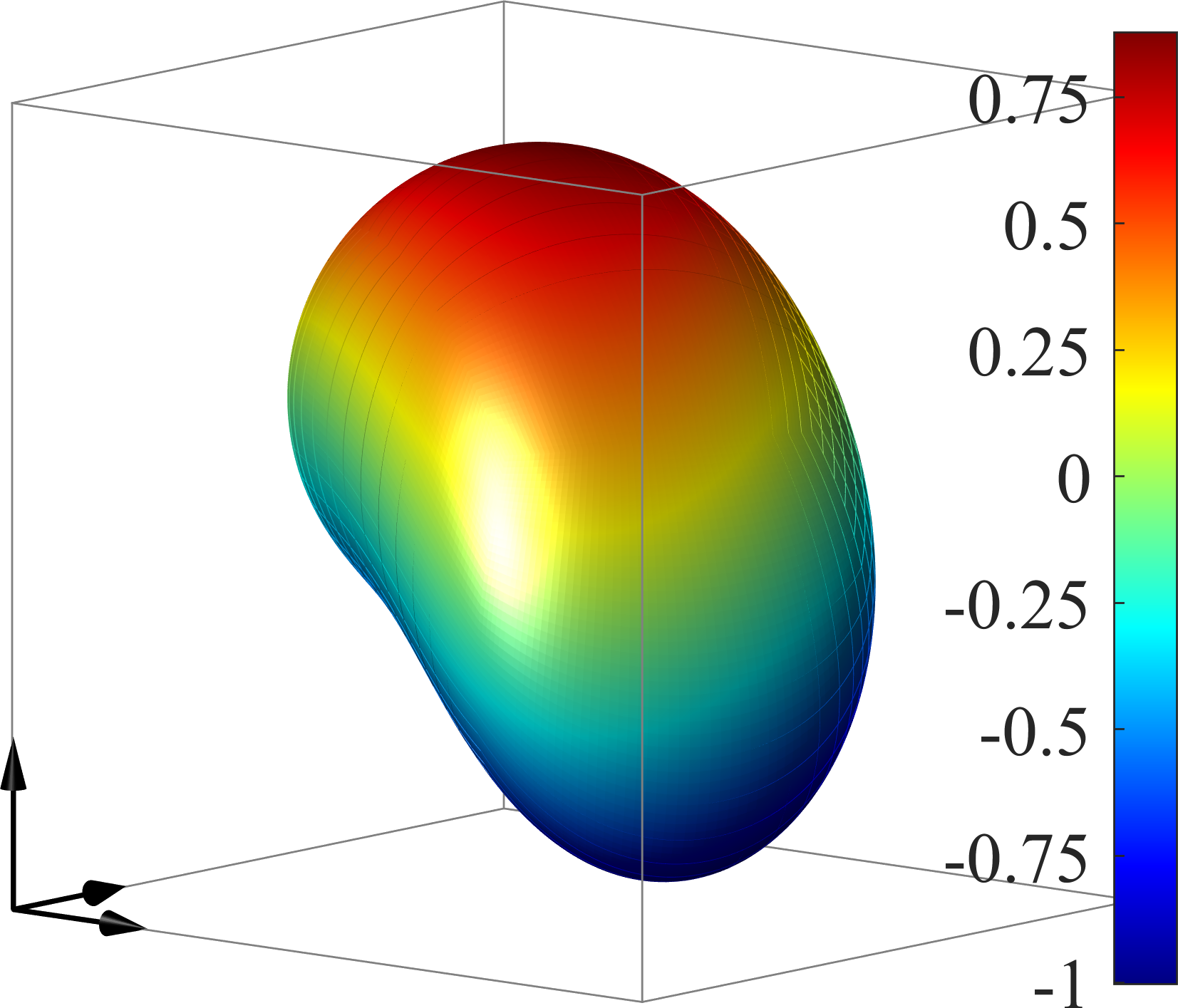}}\hfill\subfigure[$\Omega$ and level sets of $\phi$]{\includegraphics[width=0.24\textwidth]{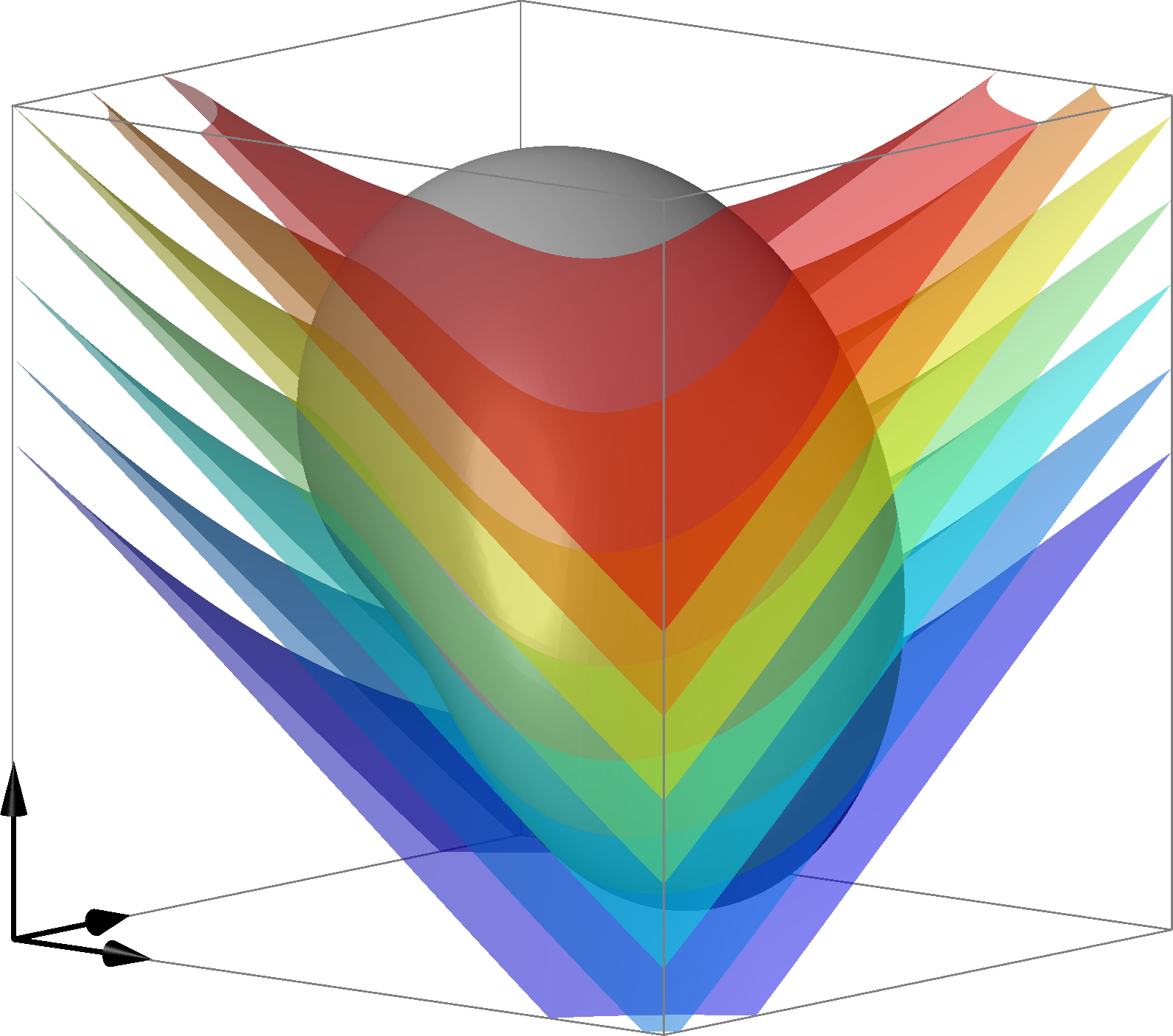}}\hfill\subfigure[$\Omega$ and level sets $\Gamma^{c}$]{\includegraphics[width=0.24\textwidth]{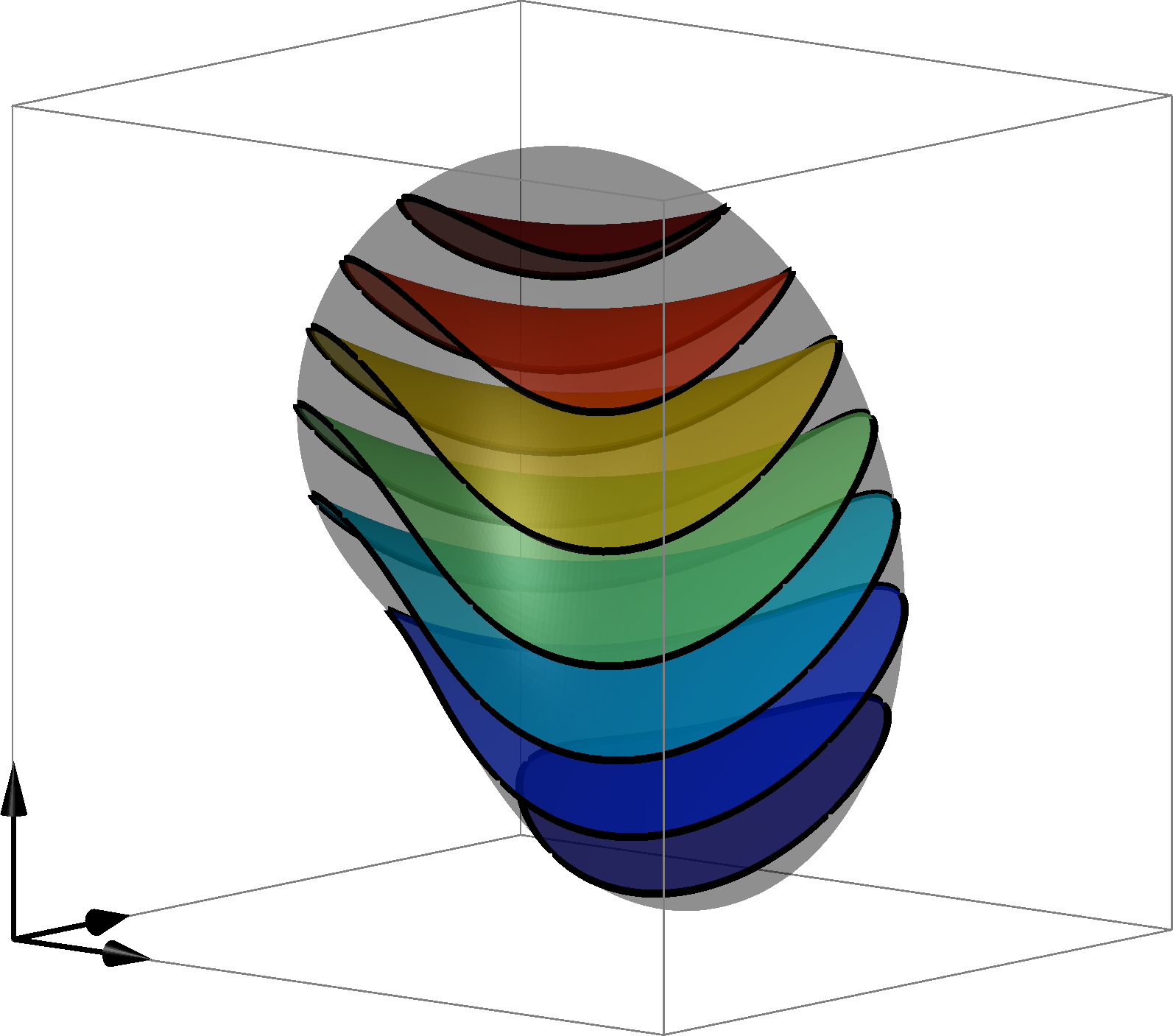}}\hfill\subfigure[level sets $\Gamma^{c}$]{\includegraphics[width=0.24\textwidth]{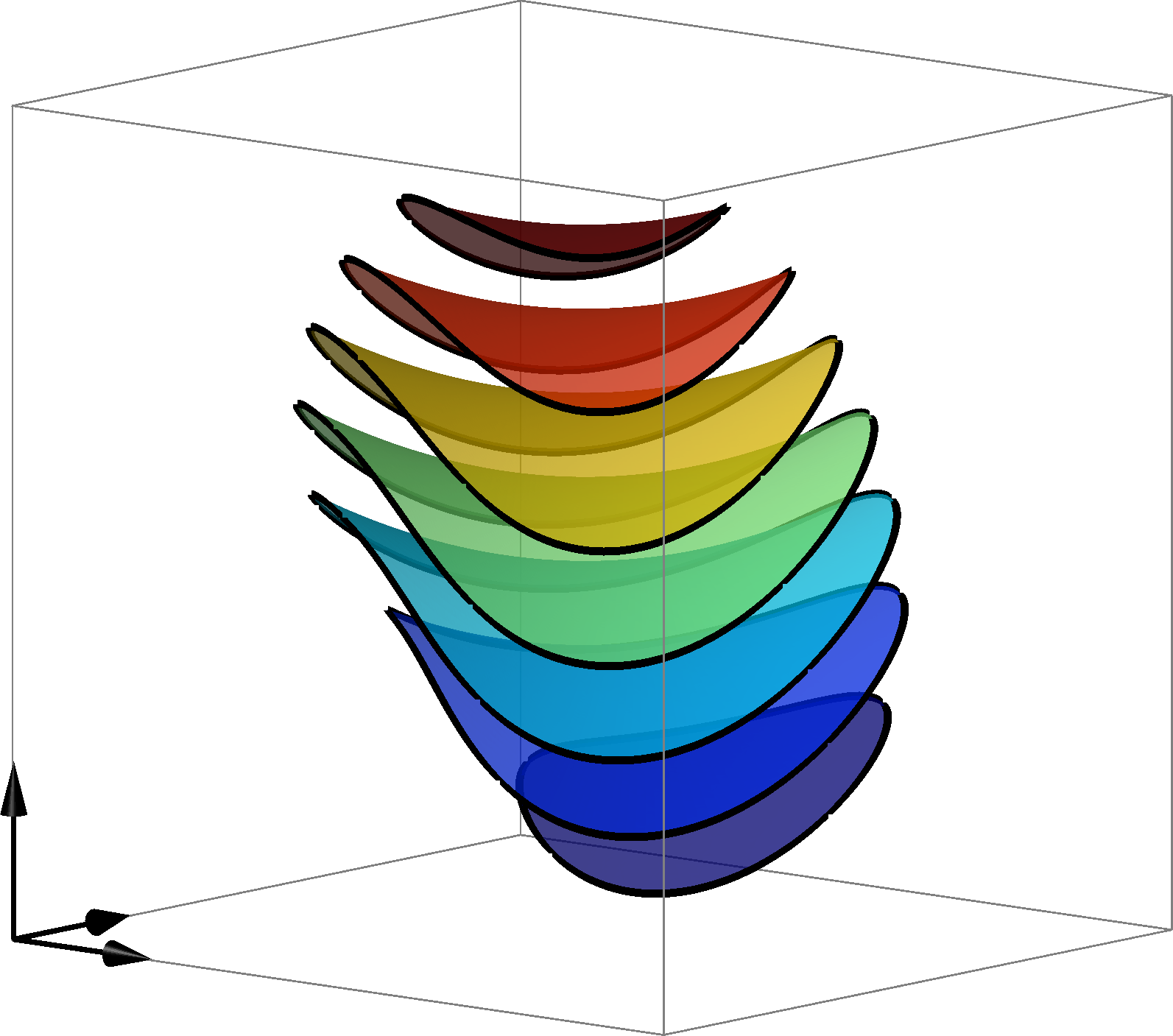}}
	
	\caption{\label{fig:BulkDomainWithLevelSets}Some bulk domain $\Omega$, level-set function $\phi\left(\vek x\right)$, and implied level sets $\Gamma^{c}$.}
\end{figure}

\begin{figure}
	\centering
	
	\subfigure[$\Omega$ and $\phi$]{\includegraphics[width=0.24\textwidth]{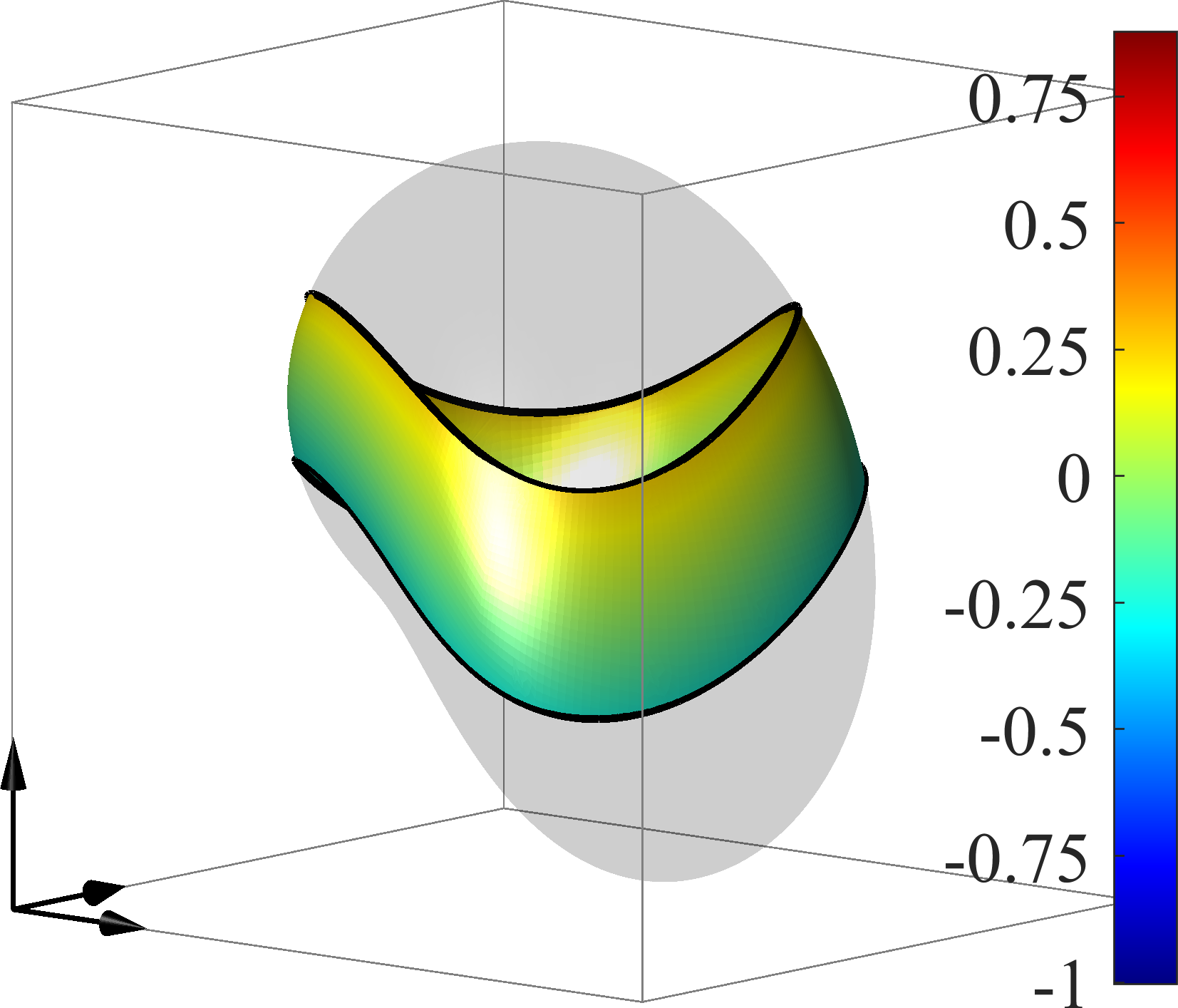}}\hfill\subfigure[$\Omega$ and level sets of $\phi$]{\includegraphics[width=0.24\textwidth]{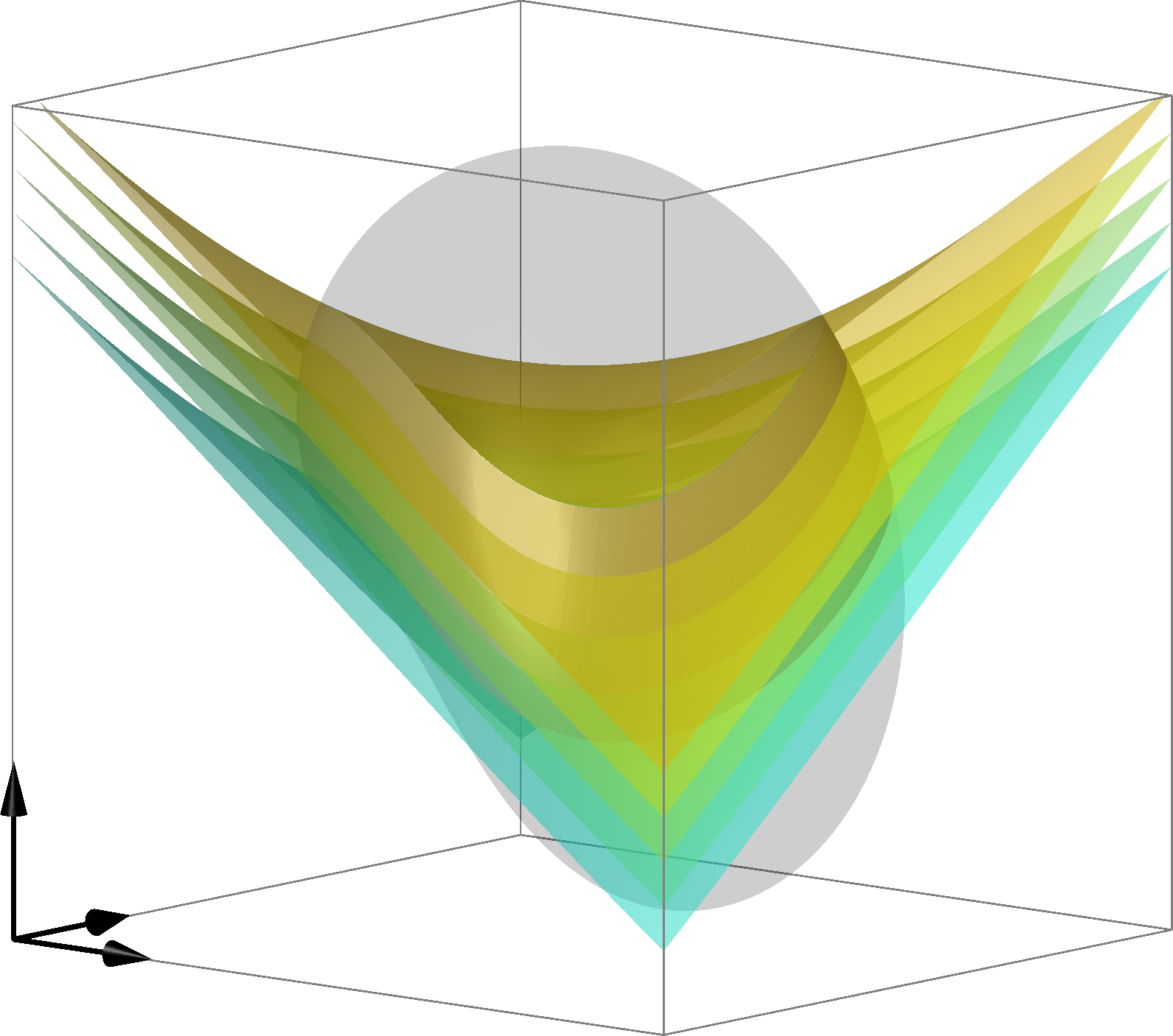}}\hfill\subfigure[$\Omega$ and level sets $\Gamma^{c}$]{\includegraphics[width=0.24\textwidth]{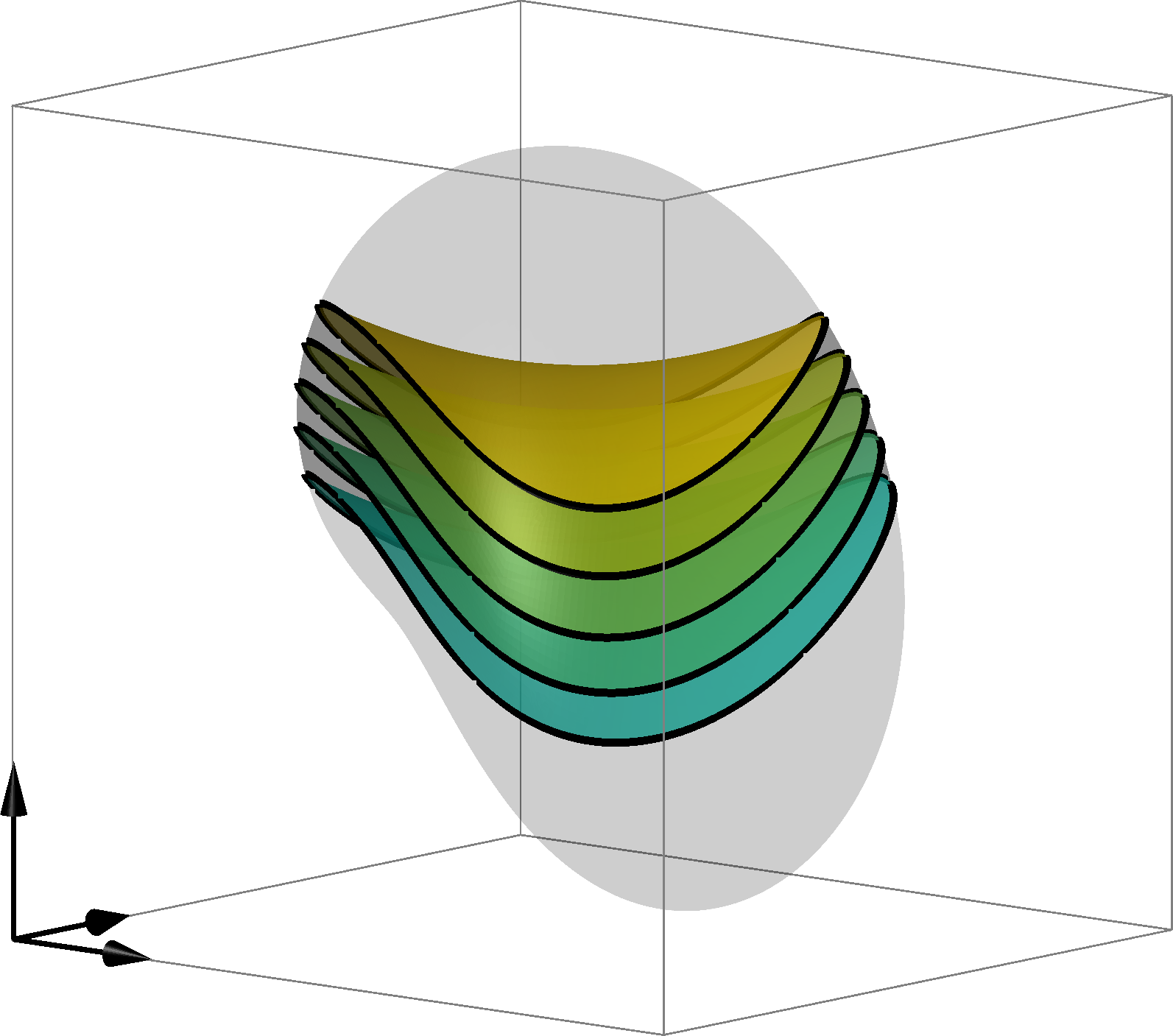}}\hfill\subfigure[level sets $\Gamma^{c}$]{\includegraphics[width=0.24\textwidth]{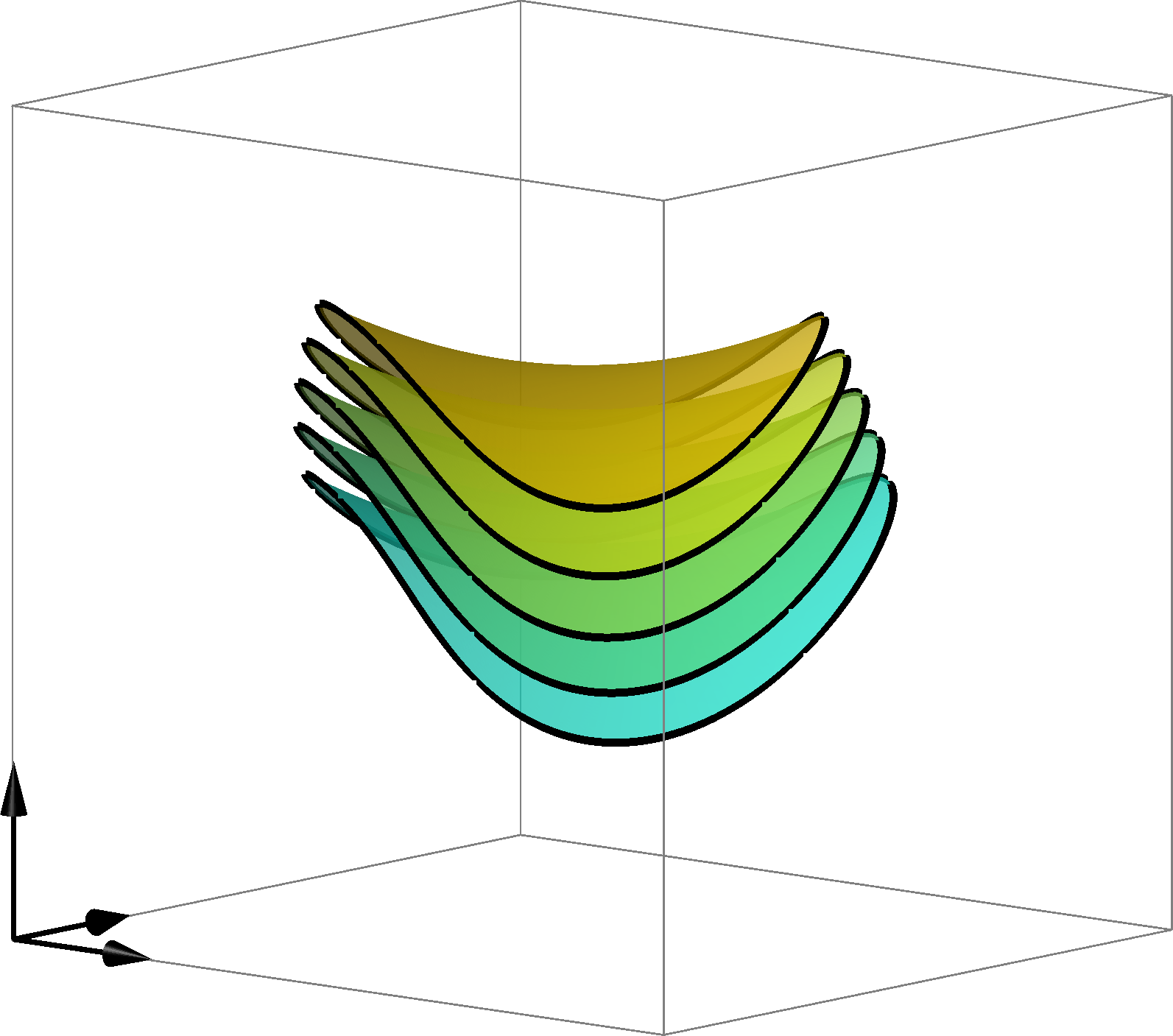}}
	
	\caption{\label{fig:BulkDomainWithLevelSetInterval}The bulk domain $\Omega$
		resulting from some larger $\Omega^{\star}$ (light gray) and a prescribed level-set interval $\left[\phi^{\min},\phi^{\max}\right]$. Some selected level sets $\Gamma^{c}$ in this interval are also shown.}
\end{figure}

Shells are curved, two-dimensional surfaces embedded in the three-dimensional physical space, hence, they are manifolds with co-dimension 1 in $\mathbb{R}^3$. Let there be a three-dimensional bulk domain $\Omega \subset \mathbb{R}^3$ and a level-set function $\phi\left(\vek x\right):\Omega\rightarrow\mathbb{R}$. Within this bulk domain exists a minimal value $\phi^{\min}=\inf\phi\left(\vek x\right)$ and a maximum value $\phi^{\max}=\sup\phi\left(\vek x\right)$ of the level-set function. The individual manifolds $\Gamma^{c}$ defined by level sets of $\phi$ with constant level-set values $c\in\mathbb{R}$, 
\begin{equation}
	\Gamma^{c}=\left\{ \vek x\in\Omega:\,\phi(\vek x)=c\in\mathbb{R}\right\},\,\phi^{\min}<c<\phi^{\max} ,\label{eq:LevelSets}
\end{equation}
are bounded, curved, two-dimensional manifolds, see Fig.~\ref{fig:BulkDomainWithLevelSets}. The boundary of some selected manifold $\Gamma^{c}$ is denoted as $\partial \Gamma^{c}$ and is the intersection curve of the level set $\Gamma^c$ with the boundary $\partial \Omega$ of the bulk domain. There holds
\begin{equation}
	\Omega = \bigcup_{c\,\in\,\Phi}\Gamma^c \text{\, and } \partial \Omega = \bigcup_{c\,\in\,\Phi} \partial \Gamma^c \text{, with } \Phi = (\phi^{\min},\,\phi^{\max}). \label{eq:setOmegaGamma}
\end{equation}
In this case, the bulk domain $\Omega$ is defined \emph{directly} with resulting $\phi^{\min}$ and $\phi^{\max}$, see again Fig.~\ref{fig:BulkDomainWithLevelSets}. Alternatively, also an \emph{indirect} definition of the bulk domain is possible: Define a superset of the bulk domain $\Omega^{\star} \! \subset \mathbb{R}^3$ and limit the bulk domain of interest $\Omega$ by specified values for $\phi^{\min}$ and $\phi^{\max}$ such that
\begin{equation}
	\Omega=\left\{ \vek x\in\Omega^{\star}:\phi^{\min}\leq\phi(\vek x)\leq\phi^{\max}\right\}. \label{eq:BulkDomainWithPrescrMinMax}
\end{equation}
This situation is depicted in Fig.~\ref{fig:BulkDomainWithLevelSetInterval}. For the proper definition of vector fields to be used in the governing equations later on, the boundary of the bulk domain $\partial \Omega$ is restricted to the parts of the boundary where $\phi\left(\vek x\right)\neq\phi^{\min}$ and $\phi\left(\vek x\right)\neq\phi^{\max}$.

\begin{figure}
	\centering
	
	\includegraphics[width=0.3\textwidth]{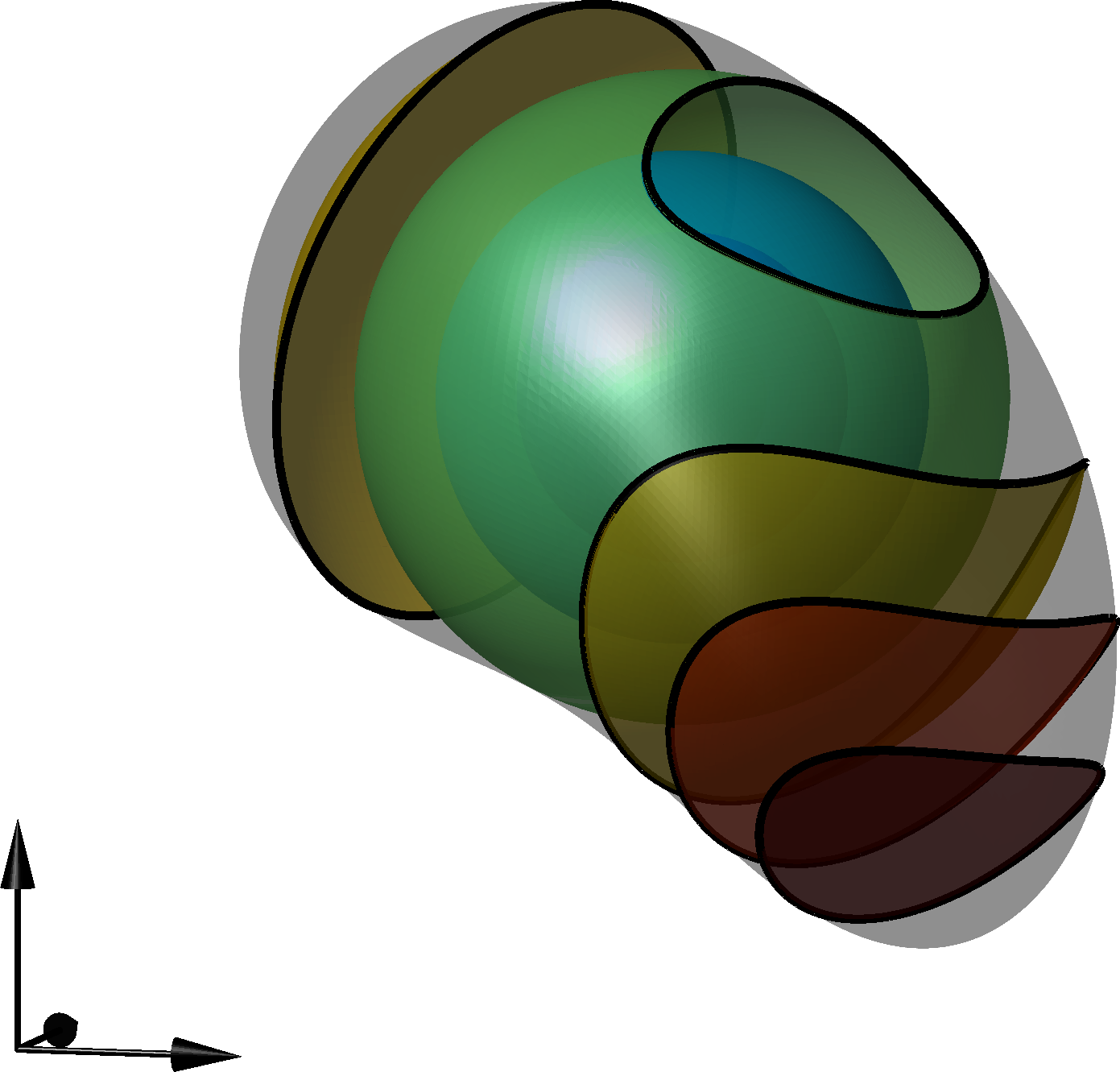}\hfill
	
	\caption{\label{fig:InvalidLevelSets} One has to ensure that the topology of the level sets varies smoothly within the bulk domain. In this example where $\phi$ features some local extremum within the bulk domain $\Omega$, some level sets are closed in $\Omega$ whereas others feature a boundary. Hence, the topology does not vary smoothly and this combination of $\phi$ and $\Omega$ is not valid.}
\end{figure}

In order to later state proper boundary value problems simultaneously on all level sets, it is crucial that within the bulk domain, the level sets vary smoothly without topology changes in the implied geometries. For example, local extreme values in the level sets may locally lead to closed level sets $(\partial \Gamma_c=\emptyset)$ whereas other level sets may feature some (smooth) boundary, see Fig.~\ref{fig:InvalidLevelSets} for a generic example. In this sense, \emph{valid combinations} of bulk domains and level-set functions which ensure smooth geometries without topological changes are required, see also \cite{Fries_2023a} for a more detailed discussion.

\subsection{Normal, co-normal and tangential vectors}\label{subsec:NormVec}

\begin{figure}
	\centering
	
	\subfigure[vectors $\vek{n}$, $\vek{m}$, $\vek{t}$, $\vek{q}$]{\includegraphics[width=0.4\textwidth]{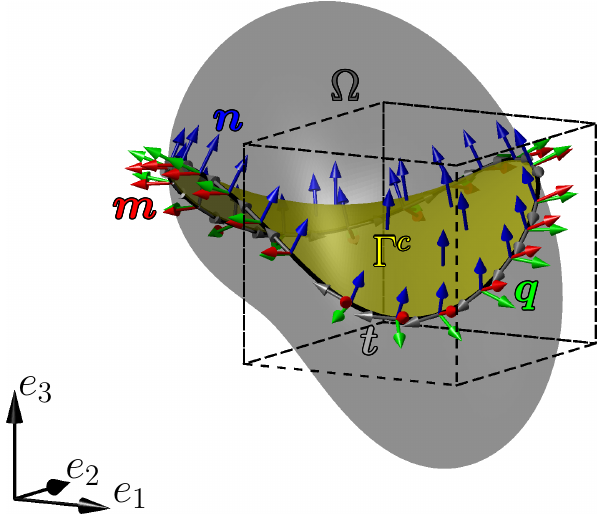}}\qquad\subfigure[zoomed view]{\includegraphics[width=0.4\textwidth]{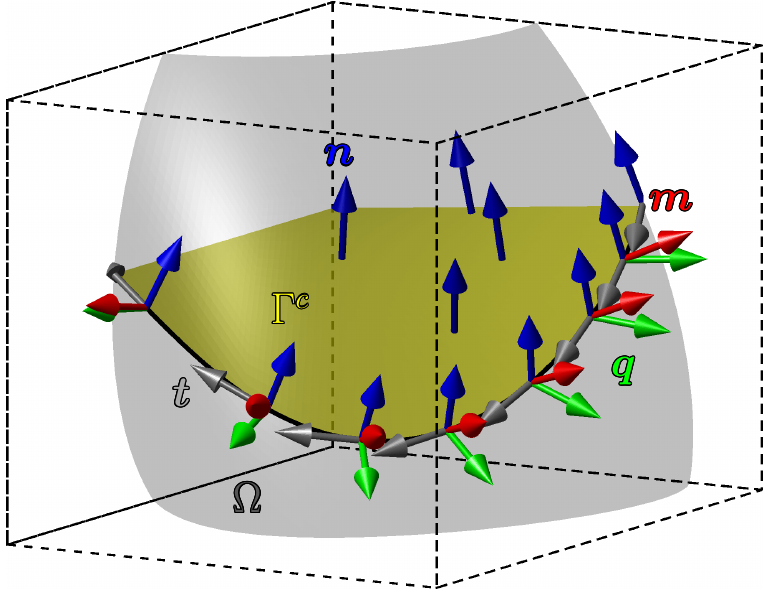}}
	
	\caption{\label{fig:VisVectors}Vector fields in the domain $\Omega$ and
		on the boundary $\partial\Omega$ shown on an arbitrarily chosen level set $\Gamma^c$ with constant $c \in \left(\phi_{\min}, \phi_{\max}\right)$. The right figure shows a zoom of the left one. Normal vectors $\vek n$ with respect to the level sets $\Gamma^{c}$ in $\Omega$ are shown in blue. Normal vectors $\vek m$ with respect to $\partial\Omega$ are red, tangential vectors $\vek t$ are gray and co-normal vectors $\vek q$ are green.}
\end{figure}

On the level sets $\Gamma^c$, the unit normal vector (field) $\vek{n}(\vek{x})$ in the whole bulk domain $\Omega$ is defined as
\begin{equation}
	\vek n\left(\vek{x}\right)=\frac{\vek{n}^{\star}}{\left\Vert \vek{n}^{\star}\right\Vert }\quad\textrm{with}\quad\vek{n}^{\star}=\nabla\phi\left(\vek{x}\right),\;\vek{x}\in\Omega,\label{eq:NormalVector_n}
\end{equation}
using the (classical) gradient of the level-set function $\phi$. In Fig.~\ref{fig:VisVectors}, the normal vector field is shown with blue arrows. Note that for brevity only one selected level set is plotted in Fig.~\ref{fig:VisVectors}. Important quantities, directly obtained from the normal vector field, are the (tangential) projector field $\mat{P}\left(\vek{x}\right) \in \mathbb{R}^{3\times 3}$ and the normal projector field $\mat{Q}\left(\vek{x}\right)\,\in \mathbb{R}^{3\times 3}$,
$\vek x\in\Omega$, 
\begin{equation}
	\mat{P}\left(\vek x\right)=\mat{I}-\vek{n}\otimes\vek{n} = \mat{I}-\mat{Q}, \qquad \mat{Q} = \vek{n}\otimes\vek{n},\label{eq:tangProjector}
\end{equation} 
where $\mat{I}$ is the $(3\times 3)$-identity matrix.\\
\\
On the boundary of the bulk domain $\partial \Omega$ lives a normal vector (field) $\vek{m}(\vek{x}),\,\vek{x}\in\partial\Omega$, which is normal to the bulk domain (instead of the level sets as for $\vek{n}$). The definition of $\vek{m}(\vek{x})$ depends on whether the bulk domain is defined explicitly or implicitly. Assuming that the bulk domain is discretized by (higher-order) elements, then computing $\vek{m}(\vek{x})$ on element boundaries is a standard operation in the FEM and, therefore, not further specified here.\\
\\
The tangent vector (field) on the boundary of the bulk domain $\partial \Omega$ is defined as
\begin{equation}
	\vek{t}\left(\vek{x}\right)= \vek{m} \times\vek{n}.\label{eq:TangVec_t}
\end{equation}
Fig.~\ref{fig:VisVectors} shows vectors $\vek{m}$ in red and tangential vectors $\vek{t}$ along the boundary in gray. A fourth vector field is defined using the normal vector of the level sets $\Gamma^c$ and the tangential vector on $\partial \Omega$, resulting in the outward unit co-normal vectors $\vek{q}(\vek{x})$
\begin{equation}
	\vek{q}\left(\vek{x}\right)=\frac{\vek{q}^{\star}}{\left\Vert \vek{q}^{\star}\right\Vert }\quad\textrm{with}\quad\vek{q}^{\star}=\vek{n} \times \vek{t}.\label{eq:Co-normalVector_q}
\end{equation}
Co-normal vectors play an important role in the formulation of the weak form of the boundary value problem (BVP) and they are fundamental parts of divergence theorems.  Fig.~\ref{fig:VisVectors} shows co-normal vectors as green arrows. It is thus seen that on the boundaries $\partial \Gamma^c$, a local triad $\left(\vek{t}, \vek{q}, \vek{n}\right)$ exists which is important to formulate boundary conditions of a Reissner--Mindlin shell later, see Section \ref{subsec:BCs}.

\subsection{Differential operators on manifolds}\label{subsec:DiffOp}
To define BVPs on manifolds, classical differential operators w.r.t. the embedding three-dimensional space and \emph{tangential} or \emph{surface} differential operators with respect to the curved, embedded, two-dimensional level sets must be distinguished. The latter are indicated by a subscript $\Gamma$, e.g., $\nabla_{\Gamma}$ for the surface gradient. We seek a formulation which does not require the introduction of a (local) curvilinear coordinate system (a parametrization) so that the resulting formulations are also applicable when the surface is defined implicitly by level sets as herein. Such a coordinate-free definition may simply be seen as a modern form of differential geometry; it is sometimes also referred to as Tangential Differential Calculus (TDC), cf.~\cite{Delfour_2011a}. The surface gradient of a scalar function $f\!\left(\vek x\right):\Omega\to\mathbb{R}$ is obtained as \cite{Delfour_2011a,Dziuk_2010a,Jankuhn_2018a,Fries_2018a}
\begin{equation}
	\nabla_{\Gamma}f=\mat{P}\cdot\nabla f,\label{eq:SurfGradScalarImplicit}
\end{equation}
where $\nabla f$ is the classical gradient in the three-dimensional
space and $\mat{P}$ the projector defined in Eq.~(\ref{eq:tangProjector}). The surface gradient of the level-set function is $\nabla_{\Gamma}\phi=\vek{0}$.

For a vector function $\vek{v}\left(\vek{x}\right):\Omega\to\mathbb{R}^{3}$, the surface gradient is applied to each component of the vector, giving the directional surface gradient of $\vek{v}\left(\vek{x}\right)$,
\begin{eqnarray}
	\nabla_{\Gamma}^{\mathrm{dir}}\vek{v} & = & \nabla\vek{v}\cdot\mat{P},\label{eq:SurfGradVectorImplicit}\\
	\text{for }\vek{v}=\left[\!\!\begin{array}{c}
		u\\
		v\\
		w
	\end{array}\!\!\right]\!\in\mathbb{R}^{3}:\left[\begin{array}{ccc}
		\partial_{x}^{\Gamma}u & \partial_{y}^{\Gamma}u & \partial_{z}^{\Gamma}u\\
		\partial_{x}^{\Gamma}v & \partial_{y}^{\Gamma}v & \partial_{z}^{\Gamma}v\\
		\partial_{x}^{\Gamma}w & \partial_{y}^{\Gamma}w & \partial_{z}^{\Gamma}w
	\end{array}\right]\! & = & \!\left[\begin{array}{ccc}
		\partial_{x}u & \partial_{y}u & \partial_{z}u\\
		\partial_{x}v & \partial_{y}v & \partial_{z}v\\
		\partial_{x}w & \partial_{y}w & \partial_{z}w
	\end{array}\right]\!\cdot\!\left[\begin{array}{ccc}
		P_{11} & P_{12} & P_{13}\\
		P_{12} & P_{22} & P_{23}\\
		P_{13} & P_{23} & P_{33}
	\end{array}\right].\nonumber 
\end{eqnarray}
For vector functions, \emph{directional} and \emph{covariant} surface gradients must be distinguished. The covariant surface gradient is defined as the projection of the directional one onto the tangent
space, 
\begin{equation}
	\nabla_{\Gamma}^{\mathrm{cov}}\vek{v}\;=\;\mat{P}\cdot\nabla_{\Gamma}^{\mathrm{dir}}\vek{v}\;=\;\mat{P}\cdot\nabla\vek{v}\cdot\mat P.\label{eq:CovariantSurfaceGradient}
\end{equation}
Note that the covariant gradient is an in-plane quantity, i.e., $ \nabla_{\Gamma}^{\mathrm{cov}}\vek{v}\in T_P\Gamma^c$, while the directional gradient is generally not in the tangent space of $\Gamma^c$, i.e., $ \nabla_{\Gamma}^{\mathrm{dir}}\vek{v}\notin T_P\Gamma^c$.\\
\\
Concerning the \emph{surface divergence }of vector functions $\vek{v}\left(\vek{x}\right) : \Gamma \to \mathbb{R}^3$
and second-order tensor functions $\mat T\negmedspace\left(\vek{x}\right):\Omega\to\mathbb{R}^{3\times 3}$,
there holds 
\begin{eqnarray}
	\mathrm{div}_{\Gamma}\,\vek{v}\left(\vek{x}\right) & = & \mathrm{tr}\left(\nabla_{\Gamma}^{\mathrm{dir}}\vek{v}\right)=\mathrm{tr}\left(\nabla_{\Gamma}^{\mathrm{cov}}\vek{v}\right)\eqqcolon\nabla_{\Gamma}\cdot\vek{v},\label{eq:DivergenceVector}\\
	\mathrm{div}_{\Gamma}\,\mat T\negmedspace\left(\vek{x}\right) & = & \left[\begin{array}{c}
		\mathrm{div}_{\Gamma}\left(T_{11},T_{12},T_{13}\right)\\
		\mathrm{div}_{\Gamma}\left(T_{21},T_{22},T_{23}\right)\\
		\mathrm{div}_{\Gamma}\left(T_{31},T_{32},T_{33}\right)
	\end{array}\right]\eqqcolon\nabla_{\Gamma}\cdot\mat T.\label{eq:DivergenceTensor}
\end{eqnarray}

\subsection{Weingarten map and curvature} \label{subsec:Curvature}

The Weingarten map \cite{Delfour_2011a,Jankuhn_2018a} is related to the second fundamental form and, thereby also to curvature. It is a symmetric, in-plane tensor which is defined as
\begin{equation}
	\mat{H} = \nabla_{\Gamma}^{\mathrm{dir}}\vek{n} = \nabla_{\Gamma}^{\mathrm{cov}}\vek{n}. \label{eq:Weingarten}
\end{equation}

The two non-zero eigenvalues are the principal curvatures, $\kappa_{1,2} = -\mathrm{eig}(\mat{H})$. The Gauß curvature is obtained as $K=\kappa_1 \cdot \kappa_2$ and the mean curvature as $\varkappa = \mathrm{tr}(\mat{H})$ \cite{Schoellhammer_2019a}.

\subsection{Integral theorems}\label{sec: IntThBTF}
For the formulation of the weak form of (partial) differential equations, which is the basis for the FEM, divergence theorems are required. A relation between integrating over all level sets $\Gamma^c$ and integrating over the bulk domain $\Omega$ is given by the \emph{co-area formula} \cite{Dziuk_2008a,Federer_1969a,Morgan_1988a,Burger_2009a,Delfour_1995a}. For an arbitrary scalar function $f\left(\vek{x}\right)$ to be integrated, the co-area formula is defined as
\begin{equation}
	\int_{\phi^{\min}}^{\phi^{\max}}\int_{\Gamma^{c}}f\left(\vek{x}\right)\;\ensuremath{\mathrm{d\ensuremath{\Gamma}}}\;\ensuremath{\mathrm{d}c}=\int_{\Omega}f\left(\vek{x}\right)\cdot\left\Vert \nabla\phi\right\Vert\;\ensuremath{\mathrm{d\ensuremath{\Omega}}}.\label{eq:CoareaFormulaDomain}
\end{equation}
For the integration over the boundary $\partial\Gamma^{c}$ in the level-set interval $\left[\phi^{\min},\;\phi^{\max}\right]$, the applied co-area formula becomes \cite{Fries_2023a}
\begin{equation}
	\int_{\phi^{\min}}^{\phi^{\max}}\int_{\partial\Gamma^{c}}f\left(\vek{x}\right)\;\ensuremath{\mathrm{d\ensuremath{\partial\Gamma}}}\;\ensuremath{\mathrm{d}c}=\int_{\partial\Omega}f\left(\vek{x}\right)\cdot\left(\vek{q}\cdot\vek{m}\right)\cdot\left\Vert \nabla\phi\right\Vert \;\ensuremath{\mathrm{d\ensuremath{\partial\Omega}}}.\label{eq:CorareaFormulaBoundary}
\end{equation}
Note that on the right hand side, the co-normal vector $\vek{q}$, defined in Eq.~(\ref{eq:Co-normalVector_q}) with respect to $\Gamma^c$, and the normal vector $\vek{m}$ on $\partial \Omega$ occur.\\
\\
The well-known divergence theorem for a vector-valued function $\vek{v}(\vek{x})$ and for a second order tensor-valued function $\mat{T}\!\left(\vek{x}\right)$ for a \emph{single} surface $\Gamma^c$, defined by \emph{one} constant value for $c$, is \cite{Schoellhammer_2019a}
\begin{equation}
	\int_{\Gamma^{c}}\vek{v}\cdot\mathrm{div}_{\Gamma}\,\mat{T}\,\mathrm{d}\Gamma=-\int_{\Gamma^{c}}\nabla_{\Gamma}^{\mathrm{dir}} \vek{v} : \mat{T}\,\mathrm{d}\Gamma+\int_{\Gamma^{c}}\varkappa\cdot \vek{v}\cdot\left(\mat{T}\cdot\vek{n}\right)\,\mathrm{d}\Gamma+\int_{\partial\Gamma^{c}}\vek{v}\cdot\left(\mat {T}\cdot\vek{q}\right)\,\mathrm{d}\partial\Gamma,\label{eq:DivTheoremVector}
\end{equation}
where $\nabla_{\Gamma}^{\mathrm{dir}} \vek{v} : \mat{T} = \mathrm{tr}\left(\nabla_{\Gamma}^{\mathrm{dir}} \vek{v} \cdot \mat{T}^{\mathrm{T}}\right)$. Note that the mean curvature $\varkappa=\mathrm{div}\,\vek{n}=\mathrm{div}_{\Gamma}\,\vek{n}$, the normal vector $\vek{n}$, and the co-normal vector $\vek{q}$ are involved on the right hand side. The middle term on the right hand side, including the mean curvature, vanishes if the tensor-valued function $\mat{T}(\vek{x})$ is in-plane, i.e., $\mat{T} = \mat{T}_t = \mat{P} \cdot \mat{T} \cdot \mat{P} \in T_P\Gamma^c$ because $\mat{T}_t \cdot \vek{n} = 0$. Combining the co-area formula,  Eqs.~(\ref{eq:CoareaFormulaDomain}) and (\ref{eq:CorareaFormulaBoundary}) with the divergence theorem Eq.~(\ref{eq:DivTheoremVector}) for \emph{one} manifold results in a divergence theorem for \emph{all} level sets in the bulk domain as

\begin{align}
	\int_{\Omega}\vek{v}\cdot\mathrm{div}_{\Gamma}\mat T\cdot\left\Vert \nabla\phi\right\Vert \,\mathrm{d}\Omega= & -\int_{\Omega}\left(\nabla_{\Gamma}^{\mathrm{dir}}\vek{v}:\mat T\right)\cdot\left\Vert \nabla\phi\right\Vert \,\mathrm{d}\Omega+\int_{\Omega}\varkappa\cdot\vek{v}\cdot\left(\mat T\cdot\vek{n}\right)\cdot\left\Vert \nabla\phi\right\Vert \,\mathrm{d}\Omega\label{eq:DivTheoremTensorBTF}\\
	& +\int_{\partial\Omega}\vek{v}\cdot\left(\mat T\cdot\vek{q}\right)\cdot\left(\vek{q}\cdot\vek{m}\right)\cdot\left\Vert \nabla\phi\right\Vert \,\mathrm{d}\partial\Omega.\nonumber 
\end{align}
This is later needed for the weak formulation in Section \ref{sec:RMS-BTF}. Note that for in-plane tensors, there holds $\nabla_{\Gamma}^{\mathrm{dir}}\vek{u}:\mat{T}_t=\nabla_{\Gamma}^{\mathrm{cov}}\vek{u}:\mat{T}_t$.

\section{Mechanical model for Reissner--Mindlin shells on all level sets}\label{sec:RMSmodel}

The strong form of \emph{one single} Reissner--Mindlin shell has been reformulated in terms of the TDC by the authors in \cite{Schoellhammer_2019b, Schoellhammer_2021a}, being applicable to one individual level set. The aim of the model formulated in this paper is to describe the mechanical behaviour of \emph{all} Reissner--Mindlin shells as implied by all level sets $\Gamma^c$ embedded in (and bounded by) the bulk domain $\Omega$,  \emph{simultaneously}. As already outlined above, the displacement and rotation fields, $\vek{u}(\vek{x})$ and  $\vek{w}(\vek{x})$, respectively, vary smoothly within the bulk domain for valid combinations of the level sets and the bulk domain, see Section \ref{subsec:GeomSetup}. To derive the strong form of the used \emph{linear} Reissner--Mindlin shell model, kinematics, equilibrium, and constitutive relations are considered as usual in structural mechanics, employing the differential operators as defined above. The equations describing this BVP look formally very similar to the ones obtained in \cite{Schoellhammer_2019b} and, in fact, if we restricted the geometric setup to one single shell geometry, these two formulations become identical.

\subsection{Kinematics} \label{subsec:kinem}

\begin{figure}
	\centering
	
	\includegraphics[width=0.6\textwidth]{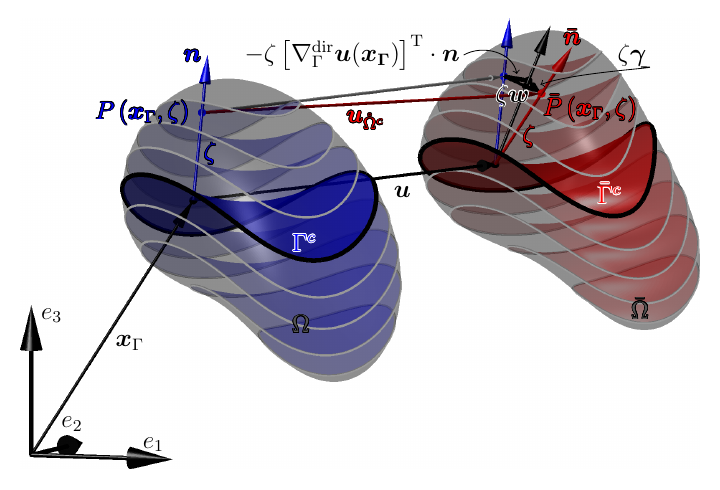}\hfill
	
	\caption{\label{fig:SF-Kin} Sketch of the kinematic relations of some Reissner--Mindlin shells embedded in the bulk domain $\Omega$ where \emph{one} middle surface $\Gamma^c$ of a shell is highlighted. The relations introduced in this section are depicted. Blue colour refers to the undeformed configuration and red colour refers to the deformed configuration.}
\end{figure}

In the Reissner--Mindlin shell model, the cross section remains straight after the deformation but not necessarily normal to the middle surface due to transverse shear deformations. This rotation of the normal vector is modelled using an in-plane difference vector, analogously to \cite{Basar_1985a,Echter_2013a,Kiendl_2017a}, therein, for a model based on curvilinear coordinates. Fig.~\ref{fig:SF-Kin} shows the kinematic setup of simultaneously considered Reissner--Mindlin shells embedded in some bulk domain. The shell continuum $\mathring{\Omega}^c$ is defined by the middle surface $\Gamma^c$ and the shell thickness $\left\vert \zeta \right\vert \leq \frac{t}{2}$ in direction of the normal vector $\vek{n}$. Note that the shell continuum $\mathring{\Omega}^c$ is not to be mixed up with the bulk domain $\Omega$. The displacement of a point $\vek{P}$ in the shell continuum $\mathring{\Omega}^c$ is obtained as the difference between the actual and the reference configuration
\begin{equation}
	\vek{u}_{\mathring{\Omega}^c}(\vek{x}) = \bar{\vek{P}}(\vek{x}) - \vek{P}(\vek{x}) \Rightarrow \vek{u}_{\mathring{\Omega}^c}(\vek{x}_{\Gamma}, \zeta) = \vek{u}(\vek{x_{\Gamma}}) + \zeta \vek{w}(\vek{x_{\Gamma}}), \label{eq:ShellDisp}
\end{equation}

with $\vek{x}_{\Gamma}$ describing a point on the middle surface, the displacement field of the middle surface $\vek{u}(\vek{x_{\Gamma}})$, and the in-plane difference vector $\vek{w}(\vek{x_{\Gamma}}) : \Gamma \rightarrow T_P \Gamma$ which describes the rotation of the normal vector $\vek{n}$. Quantities w.r.t. the deformed configuration are marked with a bar, i.e., $\bar{\bullet}$. The difference vector is defined as
\begin{equation}
	\vek{w}(\vek{x_{\Gamma}}) = - \left[\nabla_{\Gamma}^{\mathrm{dir}} \vek{u}(\vek{x_{\Gamma}})\right]^{\mathrm{T}} \cdot \vek{n} + \vek{\gamma}(\vek{x_{\Gamma}}), \label{eq:DiffVec}
\end{equation}
where the first summand describes bending of the middle surface and $\vek{\gamma}(\vek{x_{\Gamma}})$ is the transverse shear deformation \cite{Schoellhammer_2019b}.\\
\\
The linear strain tensor is given as
\begin{equation}
	\vek{\varepsilon}_{\Gamma}(\vek{x}) = \frac{1}{2} \left[ \nabla_{\Gamma}^{\mathrm{dir}} \vek{u}_{\mathring{\Omega}^c} + \left(\nabla_{\Gamma}^{\mathrm{dir}}\vek{u}_{\mathring{\Omega}^c}\right)^{\mathrm{T}} \right] = \vek{\varepsilon}_{\Gamma}^{\mathrm{P}}(\vek{x}) + \vek{\varepsilon}_{\Gamma}^{\mathrm{S}}(\vek{x}),\label{eq:LinSTgen}
\end{equation}
where an additive splitting into an in-plane strain $\vek{\varepsilon}_{\Gamma}^{\mathrm{P}}$ and a transverse shear strain $\vek{\varepsilon}_{\Gamma}^{\mathrm{S}}$ is applied. The in-plane strain is computed by
\begin{equation}
	\vek{\varepsilon}_{\Gamma}^{\mathrm{P}} = \mat{P} \cdot \vek{\varepsilon}_{\Gamma} \cdot \mat{P} = \vek{\varepsilon}_{\Gamma,\mathrm{Memb}}^{\mathrm{P}} + \zeta \vek{\varepsilon}_{\Gamma,\mathrm{Bend}}^{\mathrm{P}},\label{eq:LinST}
\end{equation}
where a split into membrane strain and bending strain is considered. Using the tangential projector $\mat{P}$, the Weingarten map $\mat{H}$, and the normal projector $\mat{Q}$, as defined in sections \ref{subsec:NormVec} and \ref{subsec:Curvature}, the membrane, bending, and transverse shear strain tensors are defined as
\begin{align}
	\vek{\varepsilon}_{\Gamma,\mathrm{Memb}}^{\mathrm{P}} (\vek{u}) & = \frac{1}{2} \left[ \nabla_{\Gamma}^{\mathrm{cov}} \vek{u} + \left(\nabla_{\Gamma}^{\mathrm{cov}}\vek{u}\right)^{\mathrm{T}} \right],\label{MembST} \\ 
	\vek{\varepsilon}_{\Gamma,\mathrm{Bend}}^{\mathrm{P}} (\vek{u},\vek{w}) & = \frac{1}{2} \left[\mat{H} \cdot \nabla_{\Gamma}^{\mathrm{dir}} \vek{u} + \left(\nabla_{\Gamma}^{\mathrm{dir}}\vek{u}\right)^{\mathrm{T}} \cdot \mat{H} + \nabla_{\Gamma}^{\mathrm{cov}} \vek{w} + \left(\nabla_{\Gamma}^{\mathrm{cov}}\vek{w}\right)^{\mathrm{T}}\right],\label{BendST} \\ 
	\vek{\varepsilon}_{\Gamma}^{\mathrm{S}} (\vek{u},\vek{w}) & = \frac{1}{2}\left[\mat{Q} \cdot \nabla_{\Gamma}^{\mathrm{dir}} \vek{u} + \left(\nabla_{\Gamma}^{\mathrm{dir}}\vek{u}\right)^{\mathrm{T}} \cdot \mat{Q} + \vek{n} \otimes \vek{w} + \vek{w} \otimes \vek{n}\right],\label{ShearST}
\end{align}
respectively, see also \cite{Schoellhammer_2019b, Schoellhammer_2021a}.

\subsection{Constitutive relations and stress resultants}\label{subsec:Constit}
A linear elastic material governed by Hooke's law and Lamé constants for plane stress, i.e., $\mu = \frac{E}{2(1+\nu)}$ and $\lambda = \frac{E\nu}{1-\nu^2}$, with Young's modulus $E$ and Poisson's ratio $\nu$, are assumed. The linear stress tensor follows as
\begin{equation}
	\vek{\sigma}_{\Gamma}(\vek{x}) = 2 \mu \vek{\varepsilon}_{\Gamma} (\vek{x}) + \lambda\,\mathrm{tr} \left[\vek{\varepsilon}_{\Gamma} (\vek{x})\right] \mat{I} \label{eq:StressTen}.
\end{equation}
This stress tensor can be decomposed into membrane and bending parts analogously to the strain tensor. A constant shifter is assumed in the material law and, therefore, an analytic pre-integration in thickness direction is possible. This leads to the definition of the stress resultants: The bending moment tensor $\mat{m}_{\Gamma}$, the effective normal force tensor $\tilde{\mat{n}}_{\Gamma}$, and the transverse shear force tensor $\mat{q}_{\Gamma}$, are defined as
\begin{align}
	\mat{m}_{\Gamma} &= \int_{-t/2}^{t/2} \zeta \, \mat{P}\cdot \vek{\sigma}_{\Gamma} \cdot \mat{P} \,\d\zeta = \frac{t^3}{12} \vek{\sigma}_{\Gamma}^{\mathrm{P}}\left(\vek{\varepsilon}_{\Gamma,\mathrm{Bend}}^{\mathrm{P}}\right), \label{eq:MT}\\
	\tilde{\mat{n}}_{\Gamma} &= \int_{-t/2}^{t/2} \mat{P}\cdot \vek{\sigma}_{\Gamma} \cdot \mat{P} \,\d\zeta = t \vek{\sigma}_{\Gamma}^{\mathrm{P}}\left(\vek{\varepsilon}_{\Gamma,\mathrm{Memb}}^{\mathrm{P}}\right), \label{eq:eNfT}\\
	\mat{q}_{\Gamma} &= \int_{-t/2}^{t/2} \mat{Q} \cdot \vek{\sigma}_{\Gamma} 
	+ \vek{\sigma}_{\Gamma} \cdot \mat{Q} \,\d\zeta = t \vek{\sigma}_{\Gamma}^{\mathrm{S}}\left(\vek{\varepsilon}_{\Gamma}^{\mathrm{S}}\right). \label{eq:SfT}
\end{align}
The moment tensor and the effective normal force tensor are in-plane, symmetric tensors. Their two non-zero eigenvalues are the principal moments and normal forces, respectively. The physical normal force tensor for curved shells is defined as $\mat{n}_{\Gamma}^{\mathrm{real}} = \tilde{\mat{n}}_{\Gamma} + \mat{H} \cdot \mat{m}_{\Gamma}$ and is in general a non-symmetric tensor. For further details, we refer to \cite{Schoellhammer_2019b}.

\subsection{Equilibrium}\label{subsec:equi-sf}
\begin{figure}
	\centering
	
	\subfigure[essential boundary conditions]{\includegraphics[width=0.4\textwidth]{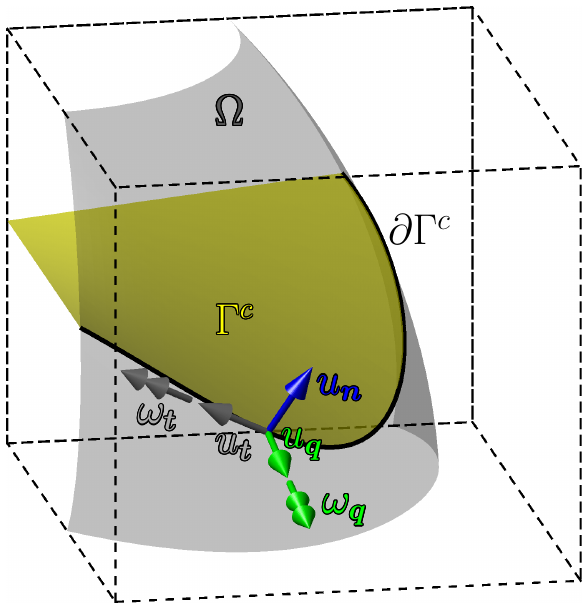}}\qquad\subfigure[natural boundary conditions]{\includegraphics[width=0.4\textwidth]{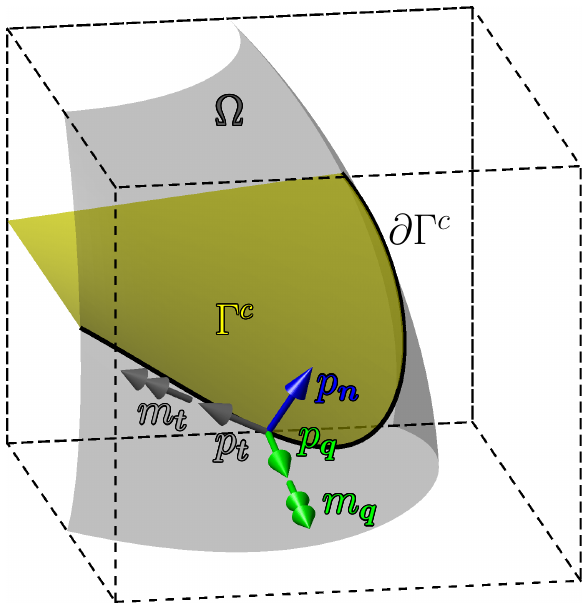}}
	
	\caption{\label{fig:VisForceMomentVectors} Displacement, rotation, force,  and moment components in terms of $\vek{t}$, $\vek{q}$, and $\vek{n}$ on the boundary $\partial\Omega$ shown on an arbitrarily chosen point at the boundary $\partial\Gamma^c$ of some level set $\Gamma^c$. Blue colour refers to quantities related to the normal vectors $\vek n$, gray refers to quantities related to the tangential vectors $\vek t$, and green refers to quantities related to the co-normal vectors $\vek q$.}
\end{figure}

The strong form of the governing equations for the linear Reissner--Mindlin shell is a system of two second-order PDEs, one resulting from the equilibrium of forces, the other from the equilibrium of moments,
\begin{align}
	\mathrm{div}_{\Gamma}\, \mat{n}_{\Gamma}^{\mathrm{real}} + \mat{Q} \cdot \mathrm{div}_{\Gamma}\, \mat{q}_{\Gamma} + \mat{H} \cdot \left(\mat{q}_{\Gamma}  \cdot \vek{n} \right) &= -\vek{f}, \label{eq:sf-force} \\
	\mat{P} \cdot \mathrm{div}_{\Gamma}\, \mat{m}_{\Gamma} - \mat{q}_{\Gamma}  \cdot \vek{n} &= -\vek{c}, \label{eq:sf-moment}
\end{align}
with $\vek{f}$ being the load vector per area, and the distributed moment vector on the middle surface $\vek{c}$. Eq.~(\ref{eq:sf-force}) can be split into an in-plane and a normal part as
\begin{align}
	\mat{P} \cdot \mathrm{div}_{\Gamma}\, \mat{n}_{\Gamma}^{\mathrm{real}} +  \mat{H} \cdot \left(\mat{q}_{\Gamma}  \cdot \vek{n} \right) &= -\vek{f}_t, \label{eq:sf-force-inpl} \\
	-\mat{H} : \mat{n}_{\Gamma}^{\mathrm{real}} + \vek{n} \cdot \mathrm{div}_{\Gamma}\, \mat{q}_{\Gamma} &= -f_n.
\end{align} 
In terms of the effective normal force, Eq.~(\ref{eq:sf-force}) is expressed as
\begin{equation}
	\mathrm{div}_{\Gamma}\, \tilde{\mat{n}}_{\Gamma} + \mat{H} \cdot \mathrm{div}_{\Gamma}\, \mat{m}_{\Gamma} + \sum_{i,j=1}^{3} \left[\mat{H}_{,i}\right]_{jk} \left[\mat{m}_{\Gamma}\right]_{ji} + \mat{Q} \cdot \mathrm{div}_{\Gamma}\, \mat{q}_{\Gamma} + \mat{H} \cdot \left(\mat{q}_{\Gamma}  \cdot \vek{n} \right) = -\vek{f}. \label{eq:sf-force-effNfT}
\end{equation}
The boundary is composed of two non-overlapping parts of $\partial \Gamma^c$, the Dirichlet boundary $\partial \Gamma^c_{\mathrm{D},i}$ and the Neumann boundary $\partial \Gamma^c_{\mathrm{N},i}$ with $i$ for $\vek{u}$ and $\vek{w}$, respectively. This leads to the following boundary conditions to be prescribed:
\begin{alignat}{3}
	\vek{u} &= \hat{\vek{g}}_{\vek{u}}  \; &\text{on} \; &\partial \Gamma_{\mathrm{D},\vek{u}}^c, \label{eq:BC-DirU}\\
	\mat{n}_{\Gamma}^{\mathrm{real}} \cdot \vek{q} + \left(\vek{n} \cdot \mat{q}_{\Gamma} \cdot \vek{q}\right) \cdot \vek{n} &= \hat{\vek{p}}_{\partial \Gamma}  \; &\text{on} \;  &\partial \Gamma_{\mathrm{N},\vek{u}}^c, \label{eq:BC-NeuU} \\
	\vek{w} &= \hat{\vek{g}}_{\vek{w}}  \; &\text{on} \; &\partial \Gamma_{\mathrm{D},\vek{w}}^c, \label{eq:BC-DirW}\\
	\mat{m}_{\Gamma} \cdot \vek{q}  &= \hat{\vek{m}}_{\partial \Gamma}  \; &\text{on} \;  &\partial \Gamma_{\mathrm{N},\vek{w}}^c. \label{eq:BC-NeuW}
\end{alignat}
Note that $\vek{q}$ is the co-normal vector at the boundary and $\mat{q}_{\Gamma}$ is the transverse shear force tensor. Fig.~\ref{fig:VisForceMomentVectors}(a) shows the essential boundary conditions for some Reissner--Mindlin shell within the bulk domain. The displacement field may be defined by the local triad $\left(\vek{t},\vek{q},\vek{n}\right)$ and with the in-plane difference vector, the rotations are defined as $\omega_{\vek{t}}=\vek{w} \cdot \vek{q}$ and $\omega_{\vek{q}}=\vek{w} \cdot \vek{t}$, respectively. For the natural boundary conditions, conjugated forces and moments can be defined as shown in Fig.~\ref{fig:VisForceMomentVectors}(b).

The BVP for the  linear Reissner--Mindlin shell on \emph{all} level sets embedded in some bulk domain is given by Eqs.~(\ref{eq:sf-force}), (\ref{eq:sf-moment}), and (\ref{eq:BC-DirU}) to (\ref{eq:BC-NeuW}). Although all embedded shells are solved simultaneously, all manifolds are perfectly decoupled, see the derivation of the weak form in section \ref{sec:RMS-BTF} and the application of the co-area formula therein (Eq. \ref{eq:CoareaFormulaDomain}).  The shell thickness $t$ is assigned to each of the manifolds describing the individual middle surface of the shells that are embedded in the bulk domain. It is assumed that all considered shells, represented by the set of individual isosurfaces, have the same thickness.

\section{The weak form for continuously embedded Reissner--Mindlin shells}\label{sec:RMS-BTF}

\subsection{Continuous weak form}\label{subsec:WF-cont}

For every FEM simulation, the governing equations have to be formulated in their weak form. For the system of PDEs considered herein, we introduce two (vector) test functions, $\vek{v}_{\vek{u}}$ for the force equilibrium (Eq.~\ref{eq:sf-force}) and $\vek{v}_{\vek{w}}$ for the moment equilibrium (Eq.~\ref{eq:sf-moment}). The corresponding function spaces are
\begin{align}
	\mathcal{S}_{\vek{u}} & =\left\lbrace \vek{v}\in\left[\mathcal{H}^{1}(\Omega)\right]^{3}:\ \vek{v}=\hat{\vek{g}}_{\vek{u}}\ \text{on}\ \partial\Omega_{\mathrm{D},\vek{u}}\right\rbrace ,\label{eq:TrialFctSpaceDispCont}\\
	\mathcal{V}_{\vek{u}} & =\left\lbrace \vek{v}\in\left[\mathcal{H}^{1}(\Omega)\right]^{3}:\ \vek v=\vek0\ \text{on}\ \partial\Omega_{\mathrm{D},\vek{u}}\right\rbrace ,\label{eq:TestFctSpaceDispCont}\\
	\mathcal{S}_{\vek{w}} & =\left\lbrace \vek {v}\in\left[\mathcal{H}^{1}(\Omega)\right]^{3}:\ \vek{v}\cdot\vek{n} = 0; \,\vek{v}=\hat{\vek{g}}_{\vek{w}}\ \text{on}\ \partial\Omega_{\mathrm{D},\vek{w}}\right\rbrace ,\label{eq:TrialFctSpaceRotCont}\\
	\mathcal{V}_{\vek{w}} & =\left\lbrace \vek{v}\in\left[\mathcal{H}^{1}(\Omega)\right]^{3}:\ \vek{v}\cdot\vek{n} = 0; \, \vek{v}=\vek{0}\ \text{on}\ \partial\Omega_{\mathrm{D},\vek{w}}\right\rbrace ,\label{eq:TestFctSpaceRotCont}
\end{align}
where $\mathcal{H}^{1}$ is the Sobolev space of functions with square
integrable first derivatives. Note that tangential function spaces are used for the test and trial functions of the difference vector $\vek{w}$ because the difference vector must be tangential due to the assumed Reissner--Mindlin kinematics, see Sections \ref{subsec:kinem} and \ref{subsec:TangConstKin}.\\
\\
To obtain the weak form, multiply the strong form of the PDE with the (vector) test functions, that is, Eq.~(\ref{eq:sf-force-effNfT}) by $\vek{v}_{\vek{u}}$ and Eq.~(\ref{eq:sf-moment}) by $\vek{v}_{\vek{w}}$, respectively, and integrate over the level sets within some interval $[\phi^{\mathrm{min}}, \phi^{\mathrm{max}}]$, $\int_{\phi^{\min}}^{\phi^{\max}}\int_{\Gamma^c}\square\,\mathrm{d} \Gamma\,\mathrm{d}c$. The co-area formula, Eq.~(\ref{eq:CoareaFormulaDomain}), converts this to an integral over the bulk domain  $\int_{\Omega}\square\cdot\left\Vert \nabla \phi\right\Vert \,\mathrm{d}\Omega$. Then, the divergence theorem, Eq.~(\ref{eq:DivTheoremTensorBTF}), is applied. For a more detailed description on how the weak form is obtained for a single shell, we refer to Section 3.5 in \cite{Schoellhammer_2019b} which applies in this context analogously.\\
\\
The weak form of the force equilibrium is now stated as: Considering given material parameters $(E,\nu) \in \mathbb{R}^{+}$, body forces $\vek{f} \in \mathbb{R}^3$ on $\Gamma^c$, tractions $\hat{\vek{p}}_{\partial\Gamma}$ on $\partial \Gamma^c_{\mathrm{N},\vek{u}}$, find $\vek{u} \in \mathcal{S}_{\vek{u}}$ and $\vek{w} \in \mathcal{S}_{\vek{w}}$ such that for all $\vek{v}_{\vek{u}} \in \mathcal{V}_{\vek{u}}$, there holds
\begin{align}
	&\int_{\Omega} \bigg[\nabla_{\Gamma}^{\mathrm{dir}} \vek{v}_{\vek{u}} : \tilde{\mat{n}}_{\Gamma} + \left(\mat{H} \cdot \nabla_{\Gamma}^{\mathrm{dir}} \vek{v}_{\vek{u}}\right) : \mat{m}_{\Gamma}  + \left(\mat{Q} \cdot \nabla_{\Gamma}^{\mathrm{dir}} \vek{v}_{\vek{u}}\right) : \mat{q}_{\Gamma}\bigg] \cdot \left\Vert \nabla \phi \right\Vert \, \mathrm{d}\Omega = \label{eq:WeakFormContForce}\\ &\int_{\Omega} \vek{v}_{\vek{u}} \cdot \vek{f} \cdot \left\Vert \nabla \phi\right\Vert \, \mathrm{d}\Omega \, + \int_{\partial \Omega_{\mathrm{N},\vek{u}}} \!\!\!\!\!\!\vek{v}_{\vek{u}} \cdot \hat{\vek{p}}_{\partial\Gamma} \cdot \left(\vek{q}\cdot\vek{m}\right) \cdot \left\Vert \nabla \phi\right\Vert \,\mathrm{d}\partial\Omega. \nonumber
\end{align}
For the moment equilibrium, the weak form is stated as: Considering given material parameters $(E,\nu) \in \mathbb{R}^{+}$, distributed moments $\vek{c} \in T_P\Gamma^c$, bending moments $\hat{\vek{m}}_{\partial\Gamma}$ on $\partial \Gamma^c_{\mathrm{N},\vek{w}}$, find $\vek{u} \in \mathcal{S}_{\vek{u}}$ and $\vek{w} \in \mathcal{S}_{\vek{w}}$ such that for all $\vek{v}_{\vek{w}} \in \mathcal{V}_{\vek{w}}$, there holds
\begin{align}
	&\int_{\Omega} \bigg(\nabla_{\Gamma}^{\mathrm{dir}} \vek{v}_{\vek{w}} : \mat{m}_{\Gamma} + \vek{v}_{\vek{w}} \cdot \mat{q}_{\Gamma} \cdot \vek{n} \bigg) \cdot \left\Vert \nabla \phi\right\Vert \, \mathrm{d}\Omega = \int_{\Omega} \vek{v}_{\vek{w}} \cdot \vek{c} \cdot \left\Vert \nabla \phi \right\Vert \, \mathrm{d}\Omega \,\, + \label{eq:WeakFormContMoment}\\ & \int_{\partial \Omega_{\mathrm{N},\vek{w}}} \!\!\!\!\!\!\vek{v}_{\vek{w}} \cdot \hat{\vek{m}}_{\partial\Gamma} \cdot \left(\vek{q}\cdot\vek{m}\right) \cdot \left\Vert \nabla \phi \right\Vert \,\mathrm{d}\partial\Omega. \nonumber
\end{align}
Note that $\mat{q}_{\Gamma}$ is the shear force tensor, $\vek{q}$ is the co-normal vector, $\mat{m}_{\Gamma}$ is the bending moment tensor, and $\vek{m}$ is the normal vector to the boundary of the bulk domain $\partial \Omega$, cf.~Sections \ref{sec:LSinBD} and \ref{sec:RMSmodel}.

\subsection{Discrete weak form} \label{sec:Impl}
Next, the weak form (Eqs.~\ref{eq:WeakFormContForce} and \ref{eq:WeakFormContMoment}) is discretized as usual in the FEM. Therefore, a conforming, three-dimensional mesh, composed by $C^0$-continuous Lagrange elements with equally spaced nodes in the reference element is generated in the bulk domain $\Omega^h$. The nodal coordinates are labelled as $\vek{x}_j$ with $j=1,\ldots,n_q$ with $n_q$ being the number of nodes in the mesh. For the test and trial functions, the resulting nodal basis functions $B_j(\vek{x})$ span a finite element space as
\begin{align}
	\mathcal{Q}_{\Omega}^{h}:=\left\lbrace v^{h}\in C^{0}(\Omega^{h}):\ v^{h}=\sum_{j=1}^{n_{q}}B_j(\vek{x})\cdot\hat{v}_{j}\text{ with }\hat{v}_{j}\in\mathbb{R}\right\rbrace \subset\mathcal{H}^{1}(\Omega^{h})\ .\label{eq:SurfaceFEMFctSpace}
\end{align}

The shape functions $B_j(\vek{x})$ are obtained by isoparametric mappings from the $3$-dimensional reference element to the physical elements. The level-set function $\phi$ is replaced by its interpolation $\phi^{h}\left(\vek{x}\right)\in\mathcal{Q}_{\Omega}^{h}$
with prescribed nodal values $\hat{\phi}_{j}=\phi\left(\vek{x}_{j}\right)$.
The following discrete test and trial function spaces, based on Eq.~(\ref{eq:SurfaceFEMFctSpace}) and Eqs.~(\ref{eq:TrialFctSpaceDispCont}) - (\ref{eq:TestFctSpaceRotCont}), are introduced 
\begin{align}
	\mathcal{S}_{\vek{u}}^h & =\left\lbrace \vek{v}^h \in\left[\mathcal{Q}_{\Omega}^{h}\right]^{3}:\ \vek{v}^h=\hat{\vek{g}}_{\vek{u}}\ \text{on}\ \partial\Omega^h_{\mathrm{D},\vek{u}}\right\rbrace ,\label{eq:TrialFctSpaceDispDiscc}\\
	\mathcal{V}_{\vek{u}}^h & =\left\lbrace \vek{v}^h \in\left[\mathcal{Q}_{\Omega}^{h}\right]^{3}:\ \vek v^h=\vek{0}\ \text{on}\ \partial\Omega^h_{\mathrm{D},\vek{u}}\right\rbrace ,\label{eq:TestFctSpaceDispDiscc}\\
	\mathcal{S}_{\vek{w}}^h & =\left\lbrace \vek {v}^h \in\left[\mathcal{Q}_{\Omega}^{h}\right]^{3}:\ \vek{v}^h\cdot\vek{n} = 0; \,\vek{v}^h=\hat{\vek{g}}_{\vek{w}}\ \text{on}\ \partial\Omega^h_{\mathrm{D},\vek{w}}\right\rbrace ,\label{eq:TrialFctSpaceRotDiscc}\\
	\mathcal{V}_{\vek{w}}^h & =\left\lbrace \vek{v}^h \in\left[\mathcal{Q}_{\Omega}^{h}\right]^{3}:\ \vek{v}^h\cdot\vek{n} = 0; \, \vek{v}^h=\vek{0}\ \text{on}\ \partial\Omega^h_{\mathrm{D},\vek{w}}\right\rbrace.\label{eq:TestFctSpaceRotDiscc}
\end{align}

The discrete weak form of Eqs.~(\ref{eq:WeakFormContForce}) and (\ref{eq:WeakFormContMoment}) follows as: Considering given material parameters $(E,\nu) \in \mathbb{R}^{+}$, body forces $\vek{f}$ and distributed moments $\vek{c}$ on $\Gamma^{c,h}$, tractions $\hat{\vek{p}}_{\partial\Gamma}$ on $\partial \Gamma^{c,h}_{\mathrm{N},\vek{u}}$ and bending moments $\hat{\vek{m}}_{\partial\Gamma}$ on $\partial \Gamma^{c,h}_{\mathrm{N},\vek{w}}$, find $\vek{u}^h \in \mathcal{S}^h_{\vek{u}}$ and $\vek{w}^h \in \mathcal{S}^h_{\vek{w}}$ such that for all test functions $\vek{v}^h_{\vek{u}} \in \mathcal{V}^h_{\vek{u}}$ and $\vek{v}^h_{\vek{w}} \in \mathcal{V}^h_{\vek{w}}$ there holds
\begin{align}
	&\int_{\Omega^h} \bigg[\nabla_{\Gamma}^{\mathrm{dir}} \vek{v}^h_{\vek{u}} : \tilde{\mat{n}}_{\Gamma} + \left(\mat{H} \cdot \nabla_{\Gamma}^{\mathrm{dir}} \vek{v}^h_{\vek{u}}\right) : \mat{m}_{\Gamma}  + \left(\mat{Q} \cdot \nabla_{\Gamma}^{\mathrm{dir}} \vek{v}^h_{\vek{u}}\right) : \mat{q}_{\Gamma}\bigg] \cdot \left\Vert \nabla \phi^h \right\Vert \, \mathrm{d}\Omega = \label{eq:WeakFormDisccForce}\\ &\int_{\Omega^h} \vek{v}^h_{\vek{u}} \cdot \vek{f} \cdot \left\Vert \nabla \phi^h\right\Vert \, \mathrm{d}\Omega \, + \int_{\partial \Omega^h_{\mathrm{N},\vek{u}}} \!\!\!\!\!\!\vek{v}^h_{\vek{u}} \cdot \hat{\vek{p}}_{\partial\Gamma} \cdot \left(\vek{q}^h\cdot\vek{m}^h\right) \cdot \left\Vert \nabla \phi^h\right\Vert \,\mathrm{d}\partial\Omega, \nonumber\\
	& \nonumber \\
	&\int_{\Omega^h} \bigg(\nabla_{\Gamma}^{\mathrm{dir}} \vek{v}^h_{\vek{w}} : \mat{m}_{\Gamma} + \vek{v}^h_{\vek{w}} \cdot \mat{q}_{\Gamma} \cdot \vek{n}^h \bigg) \cdot \left\Vert \nabla \phi^h\right\Vert \, \mathrm{d}\Omega = \int_{\Omega^h} \vek{v}^h_{\vek{w}} \cdot \vek{c} \cdot \left\Vert \nabla \phi^h \right\Vert \, \mathrm{d}\Omega \,\, + \label{eq:WeakFormDisccMoment}\\ & \int_{\partial \Omega^h_{\mathrm{N},\vek{w}}} \!\!\!\!\!\!\vek{v}^h_{\vek{w}} \cdot \hat{\vek{m}}_{\partial\Gamma} \cdot \left(\vek{q}^h\cdot\vek{m}^h\right) \cdot \left\Vert \nabla \phi^h \right\Vert \,\mathrm{d}\partial\Omega. \nonumber
\end{align}
Note that also the differential operators, e.g., $\nabla_{\Gamma} \rightarrow \nabla_{\Gamma}^h$ and the force and moment tensors, e.g., $\mat{m}_{\Gamma} \rightarrow \mat{m}_{\Gamma}^h = \mat{m}_{\Gamma}(\vek{u}^h,\vek{w}^h)$ are discretized, but we do not add the superscript $h$ to these quantities for brevity. For implementational aspects of the Bulk Trace FEM, such as the generation of customized quadrature rules and generation of surface operators applied to the FEM shape functions, see \cite{Fries_2023a}.

\subsection{A note on the difference vector}\label{subsec:TangConstKin}

As already mentioned above, the difference vector which describes the rotation of the normal vector is a tangential vector, i.e., $\vek{w} \in T_P\Gamma^c$. There are different strategies how to generate a discretized tangential vector \cite{Olshanskii_2018a,Schoellhammer_2019b}. One possibility which has been successfully used in the Bulk Trace FEM is to formulate the difference vector in the three-dimensional Cartesian coordinate system and enforcing the tangentiality constraint $\vek{w}^h \cdot \vek{n} = 0$ using a Lagrange multiplier. For the Lagrange multiplier field, an integral over the whole bulk domain is required and this leads to a significant number of additional unknowns, especially for large higher-order meshes so that we follow a different approach according to \cite{Schoellhammer_2019b}.\\
\\
Let us first assume that the difference vector is also defined in the three-dimensional Cartesian coordinate system as $\check{\vek{w}}^h = \check{w}^h_a \vek{e}_a$, according to each coordinate axis $a=\{x,y,z\}$. Then, this difference vector is projected onto the tangent space by
\begin{equation}
	\vek{w}^h = \mat{P} \cdot \check{\vek{w}}^h, \label{DiffVecDiscInPl}
\end{equation}

and the corresponding test function is obtained by $\vek{v}^h_{\vek{w}} = \mat{P} \cdot \check{\vek{v}}_{\vek{w}}^h$.

Furthermore, the gradient of the discrete, in-plane difference vector is computed using the product rule
\begin{equation}
	\nabla_{\Gamma}^{\mathrm{dir}} \vek{w}^h = \nabla_{\Gamma}^{\mathrm{dir}} \left(\mat{P} \cdot \check{\vek{w}}^h\right) = \left[\nabla_{\Gamma\,x}^{\mathrm{dir}} \,\mat{P} \cdot \check{\vek{w}}^h \,\,\, \nabla_{\Gamma\,y}^{\mathrm{dir}} \,\mat{P} \cdot \check{\vek{w}}^h \,\,\, \nabla_{\Gamma\,z}^{\mathrm{dir}} \,\mat{P} \cdot \check{\vek{w}}^h\right] + \nabla_{\Gamma}^{\mathrm{cov}} \,\check{\vek{w}}^h, \label{eq:GradDiffVec}
\end{equation}
where $\nabla_{\Gamma\,a}^{\mathrm{dir}}\,\mat{P}$ is the directional derivative of the tangential projector with respect to the coordinate axis $a$. Note that the surface derivatives of the normal vector, which are required to obtain the directional/covariant surface gradient of the tangential projector, are already part of the implementation because the Weingarten map (Eq.~\ref{eq:Weingarten})
occurs in the weak form, hence, these gradients do not cause additional computational costs. The normal part of the difference vector, $\check{w}^h_n = \check{\vek{w}}^h \cdot \vek{n}^h$, is not unique because it is not part of the discrete weak form. This would lead to an ill-conditioned system of equations, wherefore, the following stabilization term is introduced following \cite{Schoellhammer_2019b},
\begin{equation}
	s_w^h = \rho_w \int_{\Omega^h} \left(\check{\vek{w}}^h \cdot \vek{n}^h\right)\left(\check{\vek{v}}^h_w \cdot \vek{n}^h\right) \cdot \left\Vert \nabla \phi^h \right\Vert\,\mathrm{d}\Omega,
\end{equation}

which is added to the left hand side of Eq.~(\ref{eq:WeakFormDisccMoment}). The stabilization parameter is set to $\rho_w = E \cdot t$ \cite{Schoellhammer_2019b}.

\subsection{Essential boundary conditions}\label{subsec:BCs}

In the Bulk Trace FEM, prescribed Dirichlet (essential) boundary conditions can often be enforced strongly by prescribing nodal values, e.g., for the Navier support and clamped support. However, for more advanced situations at the boundary, e.g., symmetry boundary conditions including prescribed displacements and rotations, or boundary conditions in direction of a vector of the local triad at $\partial \Omega_{\mathrm{D},\vek{u}}$ and $\partial \Omega_{\mathrm{D},\vek{w}}$, these boundary conditions may have to be enforced weakly just as in single-shell analyses. Possible strategies are the Lagrange multiplier method or the symmetric/non-symmetric Nitsche's method \cite{Burman_2012a,Burman_2012b,Schillinger_2016a}. In this work, we enforce boundary conditions strongly whenever possible. Nevertheless, we shall also apply the non-symmetric Nitsche's method in the numerical example in Section \ref{subsec:SLR}, by straightforwardly adapting the approaches in \cite{Schoellhammer_2021a} for the current situation.

\section{Numerical results}\label{sec:NumRes}

For the numerical results presented here, the order of the test and trial functions for the displacement field $\vek{u}$ are \emph{one order higher} than those for the difference vector $\vek{w}$. This is along the lines of flat Reissner-Mindlin \emph{plates}, where functions related to rotations are also chosen one order lower than for bending by default.

\subsection{Error measures}\label{subsec:errorTypes}

In order to demonstrate the performance of the Bulk Trace FEM in context of the proposed mechanical model for the simultaneous solution of Reissner--Mindlin shells, different error measures are introduced next. The first two test cases are based on popular benchmark test cases for single shells and focus on the convergence towards reference displacements at selected positions. For further test cases without reference solutions given in literature, the stored energy error and the residual error are evaluated in the framework of the $h$- and $p$-version of the Bulk Trace FEM.\\
\\
For the computation of the ``residual errors'', $\varepsilon_{\mathrm{res,F}}$ for the force equilibrium and $\varepsilon_{\mathrm{res,M}}$ for the moment equilibrium, respectively, the approximated solutions $\vek{u}^h$ and $\vek{w}^h$ are inserted into the strong form, Eqs.~(\ref{eq:sf-force}) and (\ref{eq:sf-moment}). The elementwise $L_2$-error, evaluated by an integral over the elements $\Omega^{\mathrm{el},\,i}$, for the force equilibrium is obtained by
\begin{align}
	\varepsilon^2_{\mathrm{res,F}}=\; & \sum_{i=1}^{n_{\mathrm{el}}}\int_{\Omega^{\mathrm{el},\,i}}\mathfrak{r}_{\mathrm{F}}\left(\vek{u}^h,\vek{w}^h\right)\cdot\mathfrak{r}_{\mathrm{F}}\left(\vek{u}^h,\vek{w}^h\right)\cdot\left\Vert \nabla_{\!\!\vek x}\phi\right\Vert \,\mathrm{d}\Omega,\label{eq:ResidualErrorFsq}\\
	\text{with }\;\mathfrak{r}_{\mathrm{F}}\left(\vek{u}^h,\vek{w}^h\right)=\;
	&
	\mathrm{div}_{\Gamma}\,\mat{n}_{\Gamma}^{\mathrm{real}}(\vek{u}^h,\vek{w}^h) + \mat{Q} \cdot \mathrm{div}_{\Gamma}\, \mat{q}_{\Gamma}(\vek{u}^h,\vek{w}^h) + \mat{H} \cdot \left(\mat{q}_{\Gamma}(\vek{u}^h,\vek{w}^h) \cdot \vek{n} \right) + \vek{f}(\vek{x}). \nonumber		
\end{align}
For non-zero load vectors $\vek{f}$, the error values are normalized by dividing through $\int_{\Omega^{\mathrm{el},\,i}}\vek{f}^2(\vek{x}) \cdot \left\Vert \nabla_{\!\!\vek x}\phi\right\Vert\,\mathrm{d}\Omega$.

Analogously, the residual error for the moment equilibrium is obtained by
\begin{align}
	\varepsilon_{\mathrm{res,M}}^2=\; & \sum_{i=1}^{n_{\mathrm{el}}}\int_{\Omega^{\mathrm{el},\,i}}\mathfrak{r}_{\mathrm{M}}\left(\vek{u}^h,\vek{w}^h\right)\cdot\mathfrak{r}_{\mathrm{M}}\left(\vek{u}^h,\vek{w}^h\right)\cdot\left\Vert \nabla_{\!\!\vek x}\phi\right\Vert \,\mathrm{d}\Omega,\label{eq:ResidualErrorMsq}\\
	\text{with }\;\mathfrak{r}_{\mathrm{M}}\left(\vek{u}^h,\vek{w}^h\right)=\;
	&
	\mat{P} \cdot \mathrm{div}_{\Gamma}\,\mat{m}_{\Gamma}\left(\vek{u}^h,\vek{w}^h\right) - \mat{q}_{\Gamma}\left(\vek{u}^h,\vek{w}^h\right) \cdot \vek{n} + \vek{c}(\vek{x}) \nonumber.
\end{align}

For non-zero distributed moments $\vek{c}$, the error values are normalized by dividing through $\int_{\Omega^{\mathrm{el},\,i}}\vek{c}^2(\vek{x})\cdot \left\Vert \nabla_{\!\!\vek x}\phi\right\Vert\,\mathrm{d}\Omega$.

Second-order derivatives are required to evaluate the residual errors, hence, the optimal order of convergence is $\mathcal{O}(p-1)$ for FE shape functions of order $p$. Similar error measures have been used by the authors in \cite{Schoellhammer_2019a,Schoellhammer_2019b,Fries_2020a,Schoellhammer_2021a,Fries_2023a}.\\
\\
The ``stored energy error'' \cite{Zienkiewicz_2013a, Fries_2023a} is obtained by
\begin{equation}
	\varepsilon_{\mathfrak{e}}=\left|\mathfrak{e}\left(\vek{u}\right)-\mathfrak{e}\left(\vek{u}^{h}\right)\right|,\label{eq:EnergyError}
\end{equation}

with the stored (elastic) energy of the Reissner--Mindlin shell defined as
\begin{equation}
	\mathfrak{e}\left(\vek u\right) = \frac{1}{2}\int_{\Omega}
	\left[\vek{\varepsilon}_{\Gamma,\mathrm{Memb}}^{\mathrm{P}} (\vek{u}) : \tilde{\mat{n}}_{\Gamma} +
	\vek{\varepsilon}_{\Gamma,\mathrm{Bend}}^{\mathrm{P}} (\vek{u},\vek{w}) : \mat{m}_{\Gamma} +
	\vek{\varepsilon}_{\Gamma}^{\mathrm{S}} (\vek{u},\vek{w}) : \mat{q}_{\Gamma}\right] \cdot \lVert \nabla \phi \rVert
	\,\mathrm{d}\Omega.\label{eq:EnergyDef}
\end{equation}

The expected order of convergence is $\mathcal{O}(p+1)$ for element order $p$. Note that this error measure is not to be mixed with the (classical) energy error norm, cf. \cite{Zienkiewicz_2013a}.

\subsection{Scordelis-Lo roof}\label{subsec:SLR}

\begin{figure}
\begin{minipage}{0.5\textwidth}
	\includegraphics[width=\textwidth]{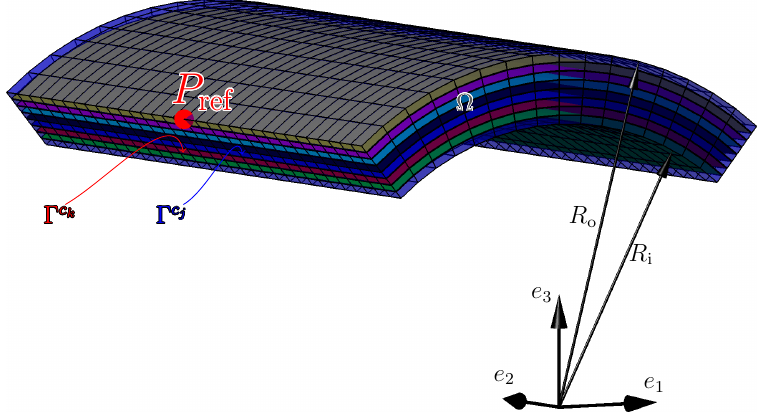}
\end{minipage}
\hspace{0.2cm}
\begin{minipage}{0.5\textwidth}
	{\footnotesize
	\begin{tabular}{l l}
		Geometry: & cylindrical shells embedded\\
		& in a cylindrical bulk domain \\
		& $L_y = 50$ \\
		& $R_\mathrm{o} = 25$ \\
		& $R_\mathrm{i} = 20$ \\
		& $\Theta = 80^{\circ}$ \\
		& $t = 0.25$ \\
		Material parameters: & $E = 4.32 \cdot 10^8$ \\
		& $\nu = 0.0$ \\ 
		& $\alpha_s = 1.0$ \\ 
		Loading: & $\vek{f} = [0,0,-90]^{\mathrm{T}}$ \\
		& $\vek{c} = \vek{0}$  \\
		Support: & rigid diaphragms at\\
		& the ends of the shells
	\end{tabular}}
\end{minipage}
\caption{\label{fig:SLR-Definition}Geometry of Scordelis-Lo roof and other, neighbouring cylindrical shells embedded in the bulk domain and definition of the parameters for these shells. The maximum vertical displacement at $P_\mathrm{ref}$ is considered in the convergence studies. $\Gamma^{c_j}$ and $\Gamma^{c_k}$ indicate two examples of middle surfaces of considered shells for different values $j,k \in c$.}
\end{figure}

\begin{figure}
	\centering
	
	\includegraphics[width=0.6\textwidth]{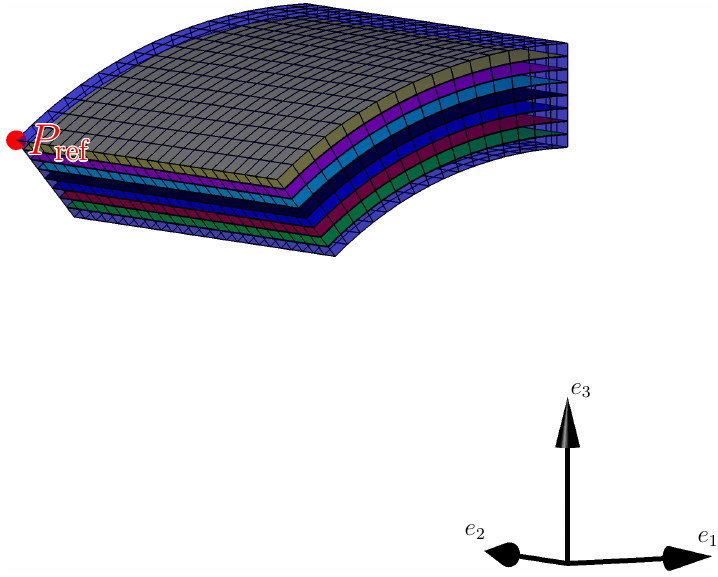}\hfill
	
	\caption{\label{fig:SRLsymmBCs} Geometry definition of the Bulk Trace FEM test case, using the well known Scordelis-Lo roof as a benchmark example, for the application of symmetry boundary conditions. Here the lower left quarter of the geometry is considered to minimize the number of degrees of freedom. The reference point for the displacement $P_{\mathrm{ref}}$ is now located in one of the corners.}
\end{figure}

\begin{figure}
	\centering
	
	\subfigure[dispalcements]{\includegraphics[width=0.42\textwidth]{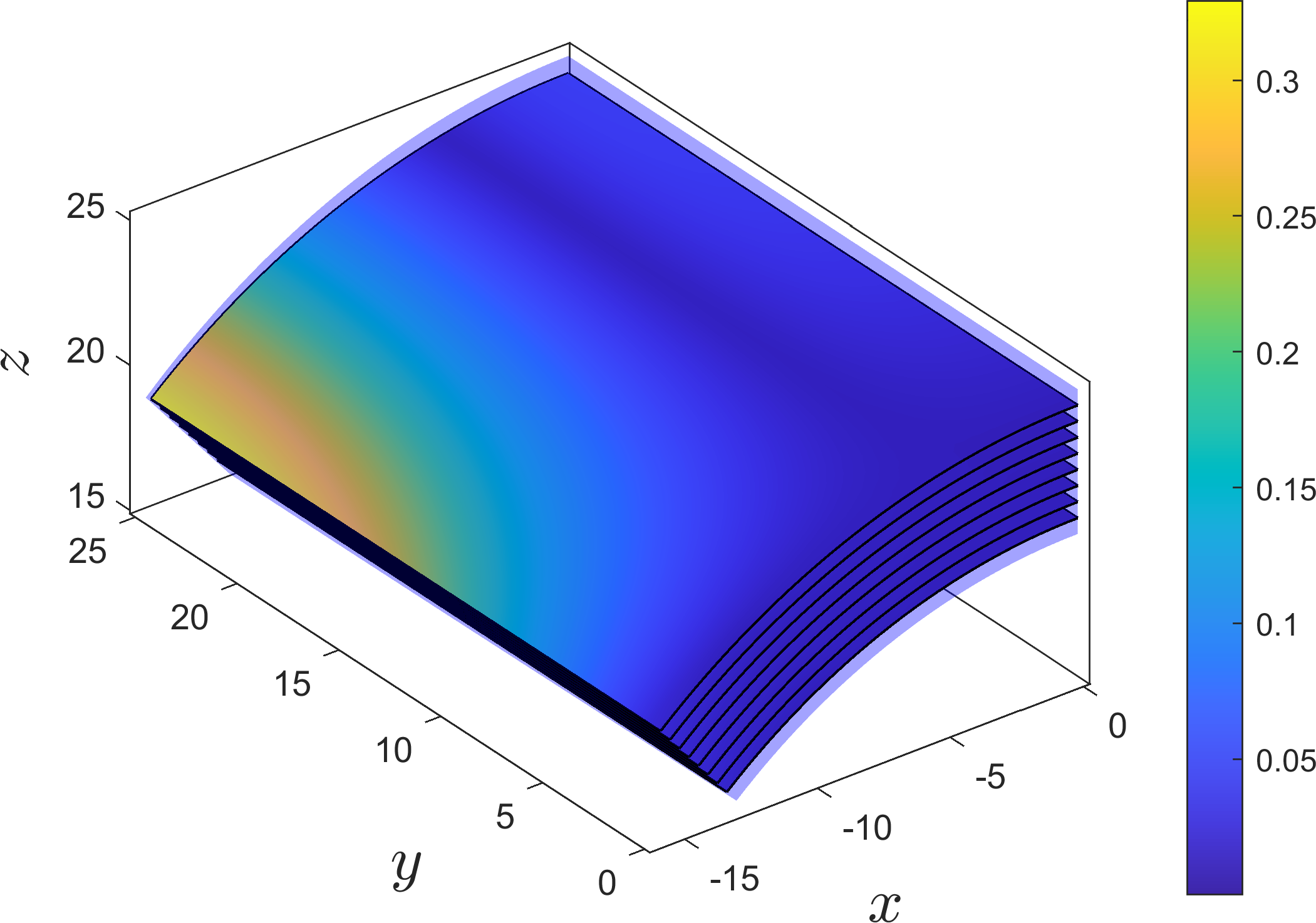}}\qquad\subfigure[normalized convergence]{\includegraphics[width=0.4\textwidth]{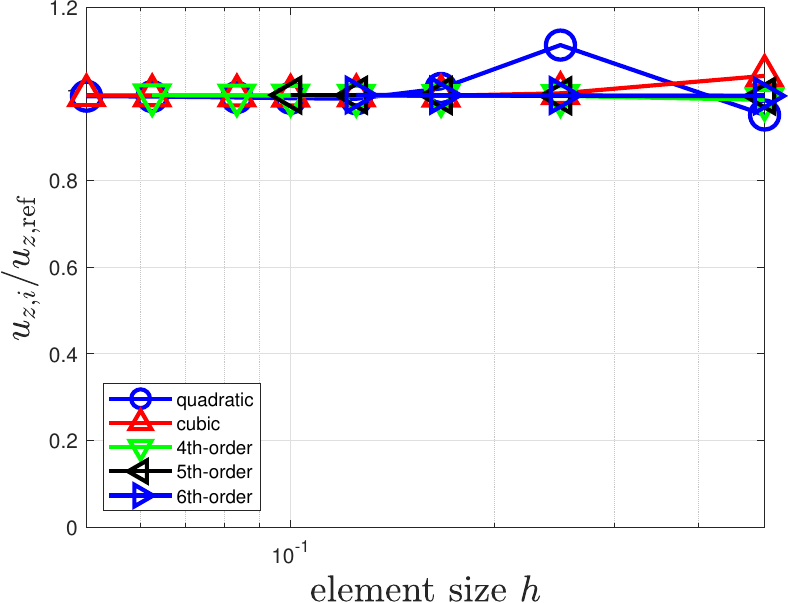}}
	
	\caption{\label{fig:ConvSLR} (a) One quarter of the bulk domain including the Scordelis-Lo roof with symmetry boundary conditions considered in the numerical analysis. The colors show the Euclidean norm of the displacements $\vek{u}$. (b) Normalized convergence of the reference displacement $u_{z,\mathrm{max,ref}}=-0.3024$.}
\end{figure}

\begin{figure}
	\centering
	
	\subfigure[ $m_{11}$]{\includegraphics[width=0.3\textwidth]{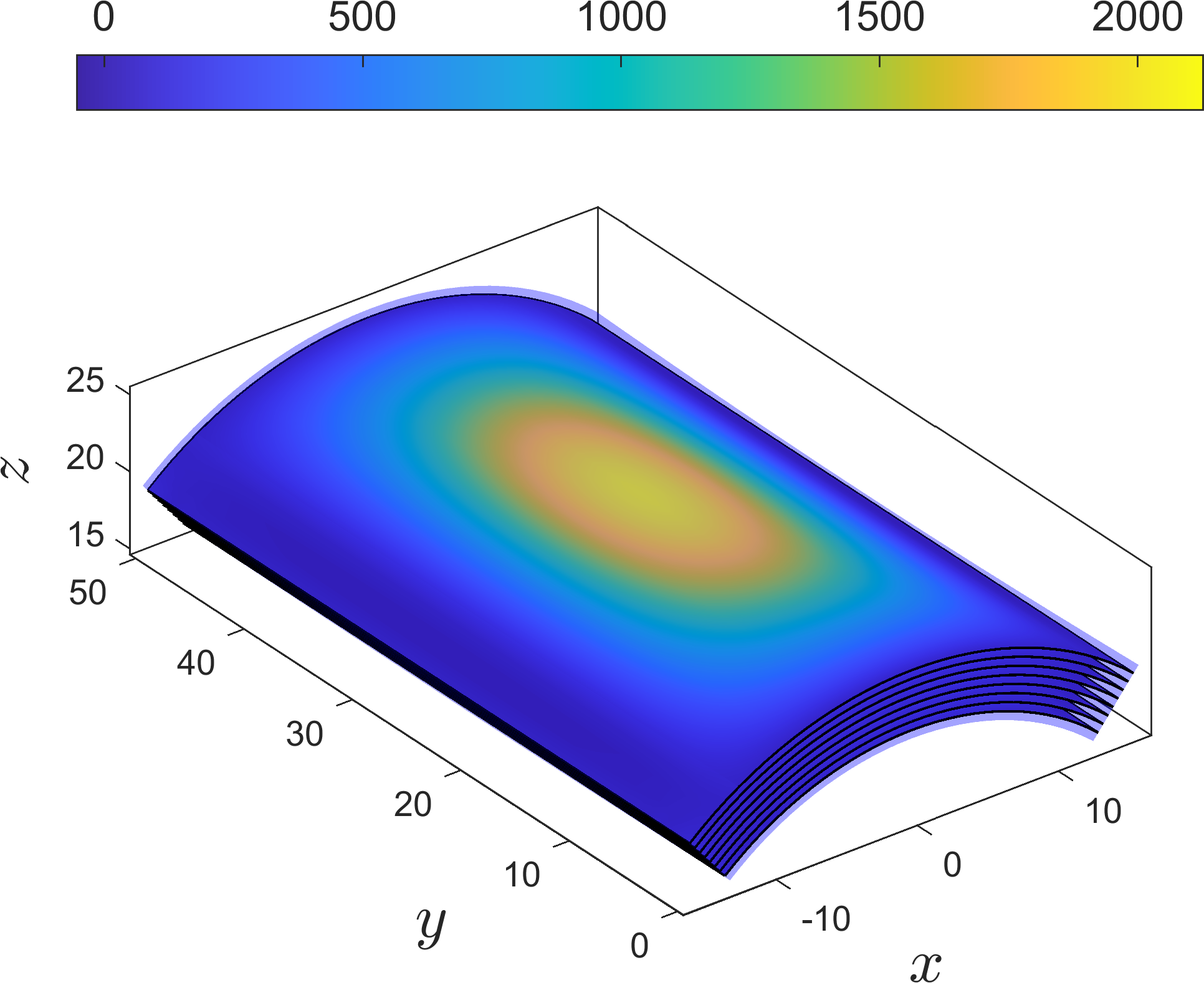}}\hfill\subfigure[ $m_{12}$]{\includegraphics[width=0.3\textwidth]{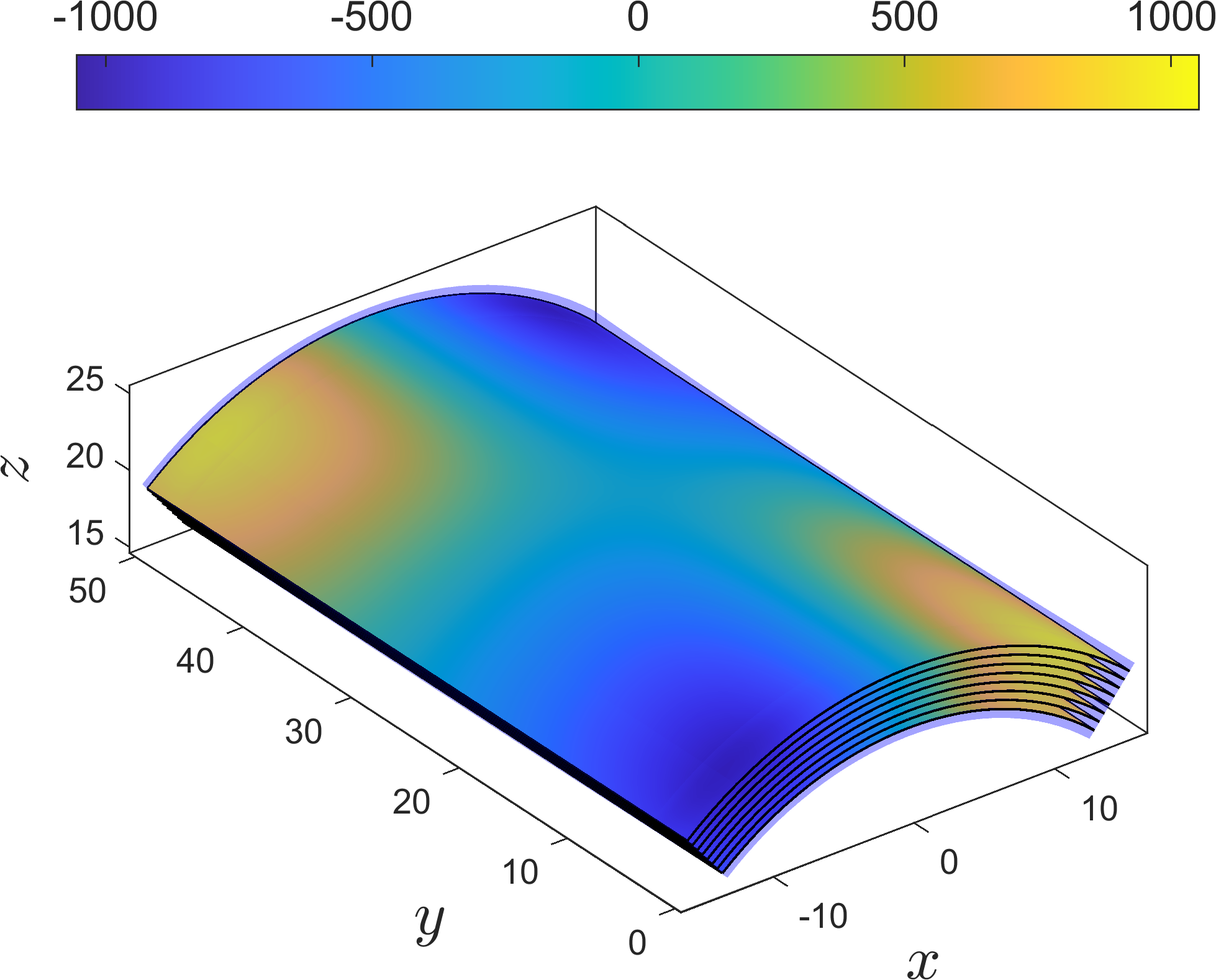}}\hfill\subfigure[$m_{22}$]{\includegraphics[width=0.3\textwidth]{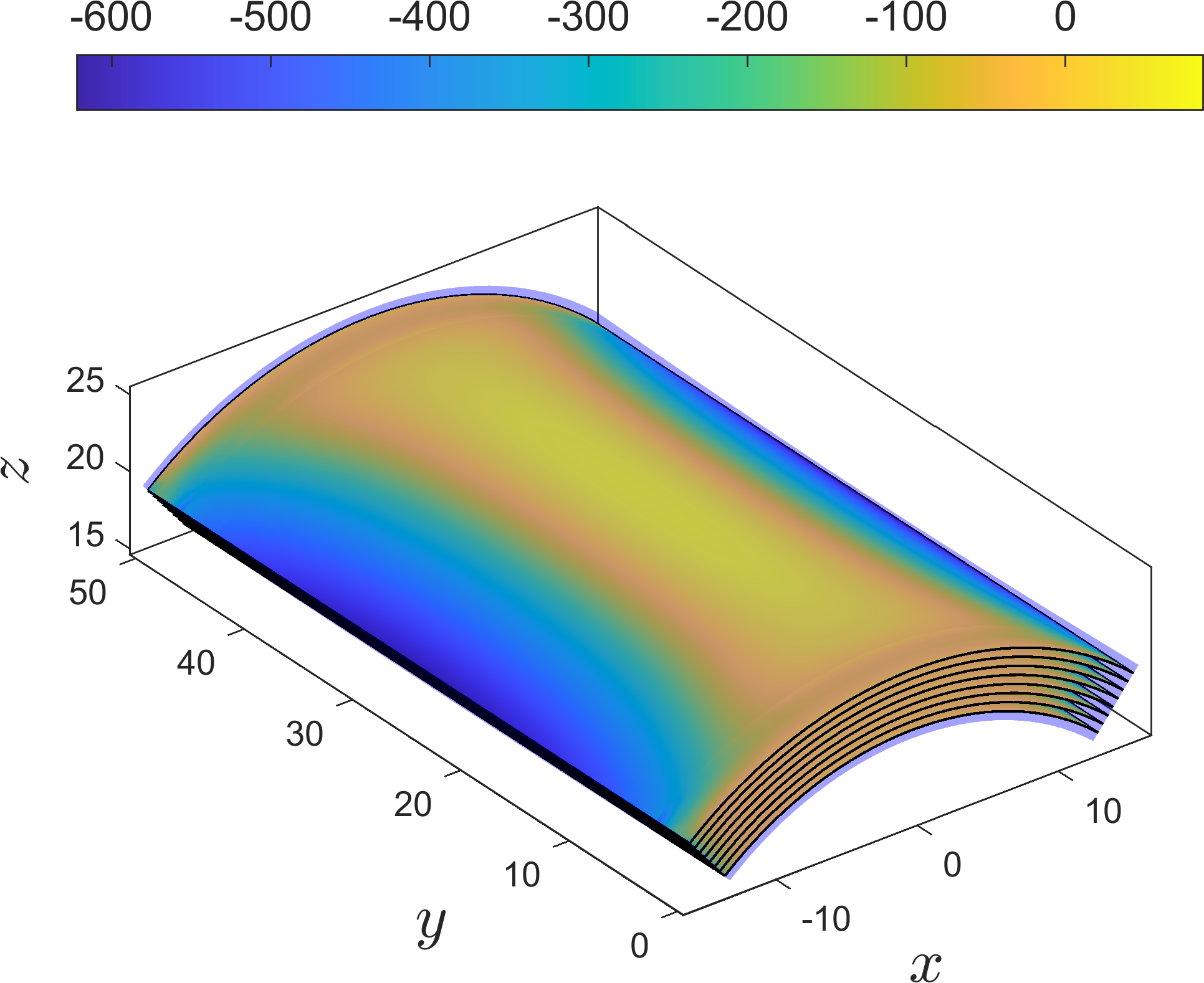}}\hfill
	
	\caption{\label{fig:SLR-BM} Components $m_{11}$, $m_{12}$, and $m_{22}$ of the bending moment tensor depicted on some selected level sets. The boundary surface (envelop) of the bulk domain, which contains the Scordelis-Lo roof and geometrically related cylindrical shells, is depicted in light blue.}
\end{figure}

The Scordelis-Lo roof, as shown in Fig.~\ref{fig:SLR-Definition}, is part of the famous \emph{shell obstacle course} \cite{Belytschko_1985a} and was often used in the literature to verify scientific contributions in classical shell mechanics, e.g., \cite{Schoellhammer_2019a,Echter_2013a,Schoellhammer_2019b,Kiendl_2017a}. The reference solution given in \cite{Belytschko_1985a} for the deformation of the middle surface of the shell at the points $\left[\pm R\cdot\cos(50^{\circ}),25,R\cdot\sin(50^{\circ})\right]^{\mathrm{T}}$ is given as $u_{z,\mathrm{max,ref}}=-0.3024$. In the present example, the shell that coincides with the \emph{single} shell from the shell obstacle course is the one which coincides with the outer layer of the bulk domain interval. The geometric setup and mechanical parameters of this test case are given in Fig.~\ref{fig:SLR-Definition}. Considering only a quarter of the shell geometry for the numerical analysis is possible due to the symmetry of the problem. However, a weak enforcement of the boundary conditions becomes necessary in this situation, see Section \ref{subsec:BCs}. The geometric setup is shown in Fig.~\ref{fig:SRLsymmBCs} and the material parameters are given next to Fig.~\ref{fig:SLR-Definition}. In Figs.~\ref{fig:SLR-Definition} and \ref{fig:SRLsymmBCs}, some (arbitrarily selected) level sets describing certain shell geometries are shown in different colours. The depicted mesh is an arbitrary example from the series of different resolutions $h$ and element orders $p$ used in the analyses. The convergence analysis in Fig.~\ref{fig:ConvSLR} clearly shows that the expected reference value for the displacements is well approximated, especially for higher-order elements, even for the coarsest meshes studied.\\
\\
A further confirmation of the proposed model and its approximation by the Bulk Trace FEM can be seen in the bending moments of the Scordelis-Lo roof. These are given in \cite{Kiendl_2017a} for a \emph{single} shell with the same geometry and material parameters as the Scordelis-Lo roof proposed in \cite{Belytschko_1985a}. Fig.~\ref{fig:SLR-BM} shows the bending moments on some level sets embedded in the bulk domain as defined in Fig.~\ref{fig:SLR-Definition}. Comparison shows that the results of the bending moments agree very well with those published in \cite{Kiendl_2017a}. 

\subsection{Partly clamped hyperbolic paraboloid}\label{subsec:HCP}

\begin{figure}
	\begin{minipage}{0.5\textwidth}
		\includegraphics[width=\textwidth]{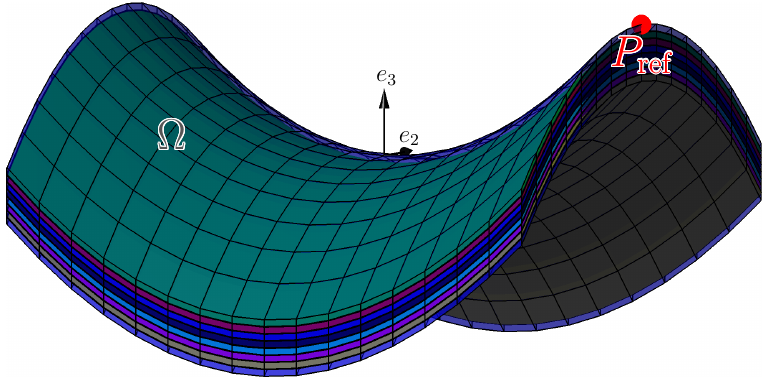}
	\end{minipage}
	\hspace{0.2cm}
	\begin{minipage}{0.5\textwidth}
		{\footnotesize
			\begin{tabular}{l l}
				Geometry: & hyperbolic paraboloid shells \\
				& $L_x = L_y = 1$ \\
				& shell: $z = x^2-y^2, \, \, t = 0.01$ \\
				& bulk domain: $h = 0.1$ \\
				& $2d$-shell surface extruded along $z$-axis\\
				Material parameters: & $E = 2.0\cdot 10^{11}$ \\
				& $\nu = 0.3$ \\ 
				& $\alpha_s = 1.0$ \\ 
				Loading: & $\vek{f} = [0,0,-8000 \cdot t]^{\mathrm{T}}$ \\
				& $\vek{c} = \vek{0}$  \\
				Support: & clamped at edge $x = -0.5$
		\end{tabular}}
	\end{minipage}
	\caption{\label{fig:CHP-Definition} Geometry of the bulk domain with some embedded clamped hyperbolic paraboloid shells and definition of the parameters for these shells. The red dot indicates the reference point $P_\mathrm{ref}$ for the benchmark displacement.}
\end{figure}

\begin{figure}
	\centering
	
	\subfigure[dispalcements]{\includegraphics[width=0.45\textwidth]{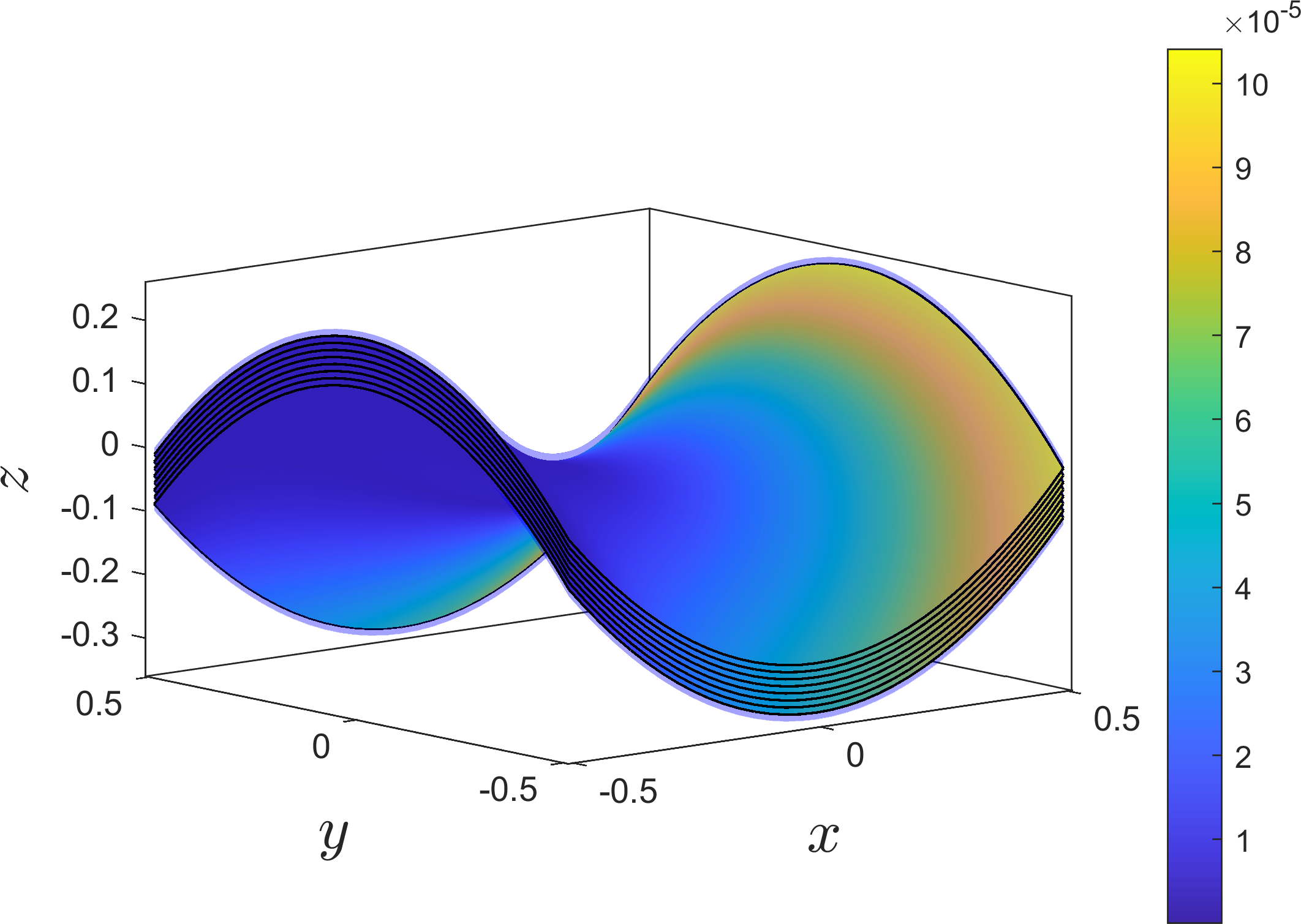}}\qquad\subfigure[normalized convergence]{\includegraphics[width=0.4\textwidth]{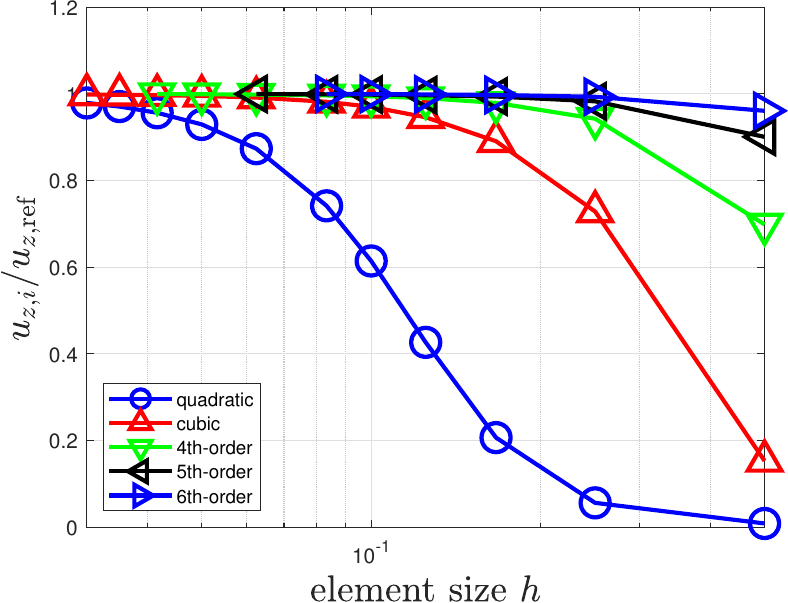}}
	
	\caption{\label{fig:ConvPCH} (a) The Euclidean norm of the displacement field for the partly clamped hyperboloids on some selected level sets. (b) Normalized convergence of the reference displacement $u_{z,\mathrm{ref}}=-9.3355\cdot 10^{-5}$.}
\end{figure}

The partly clamped hyperbolic paraboloid from \cite{Chapelle_1998a} has been used in various publications in shell mechanics to verify the results, e.g., \cite{Bathe_2000a,Schoellhammer_2019b,Schoellhammer_2021a,Kiendl_2017a}. The geometry and material parameters are shown in Fig.~\ref{fig:CHP-Definition}. In this example, the bulk domain is composed of the geometry of one shell extruded along the $z$-axis with a height $h=0.1$. The reference displacement at $P_{\mathrm{ref}} = \left[0.5, 0, 0.5\right]^{\mathrm{T}}$, marked by a red bullet point in Fig.~\ref{fig:CHP-Definition}, is $u_{z,\mathrm{ref}} = -9.3355\cdot10^{-5}$. Fig.~\ref{fig:ConvPCH}(b) shows that the numerical results converge to the reference displacement, while higher-order elements converge faster. The convergence behaviour is in agreement with previous results, e.g., \cite{Schoellhammer_2019b}. This widely used benchmark test case shows again that the proposed mechanical model and numerical method lead to the expected result. For lower element orders, mild locking phenomena are observed for coarser meshes. This behaviour is well in accordance to other results, e.g., \cite{Schoellhammer_2019b,Schoellhammer_2021a,Kiendl_2017a,Bathe_2000a}.  

\subsection{Generalized test case with Navier support}\label{subsec:iTCns}

The first two numerical examples are adaptions of well known benchmark test cases which are intended to confirm the validity of the newly introduced mechanical model and numerical method based on proven, accepted, and widely used examples for single shell geometries. The following examples have a more general geometry and are designed such that smooth solutions of (all) physical fields result and, consequently, higher-order convergence rates are expected.\\
\\
In the first of these examples, the geometry of the shells is described by isosurfaces of the level-set function
\begin{equation}
	\phi(\vek{x}) = z - 2 \cdot \sin \left(\frac{1}{4} \cdot x \cdot y\right).
\end{equation}

\begin{figure}
	\centering
	
	\subfigure[level sets and bulk domain]{\includegraphics[width=0.3\textwidth]{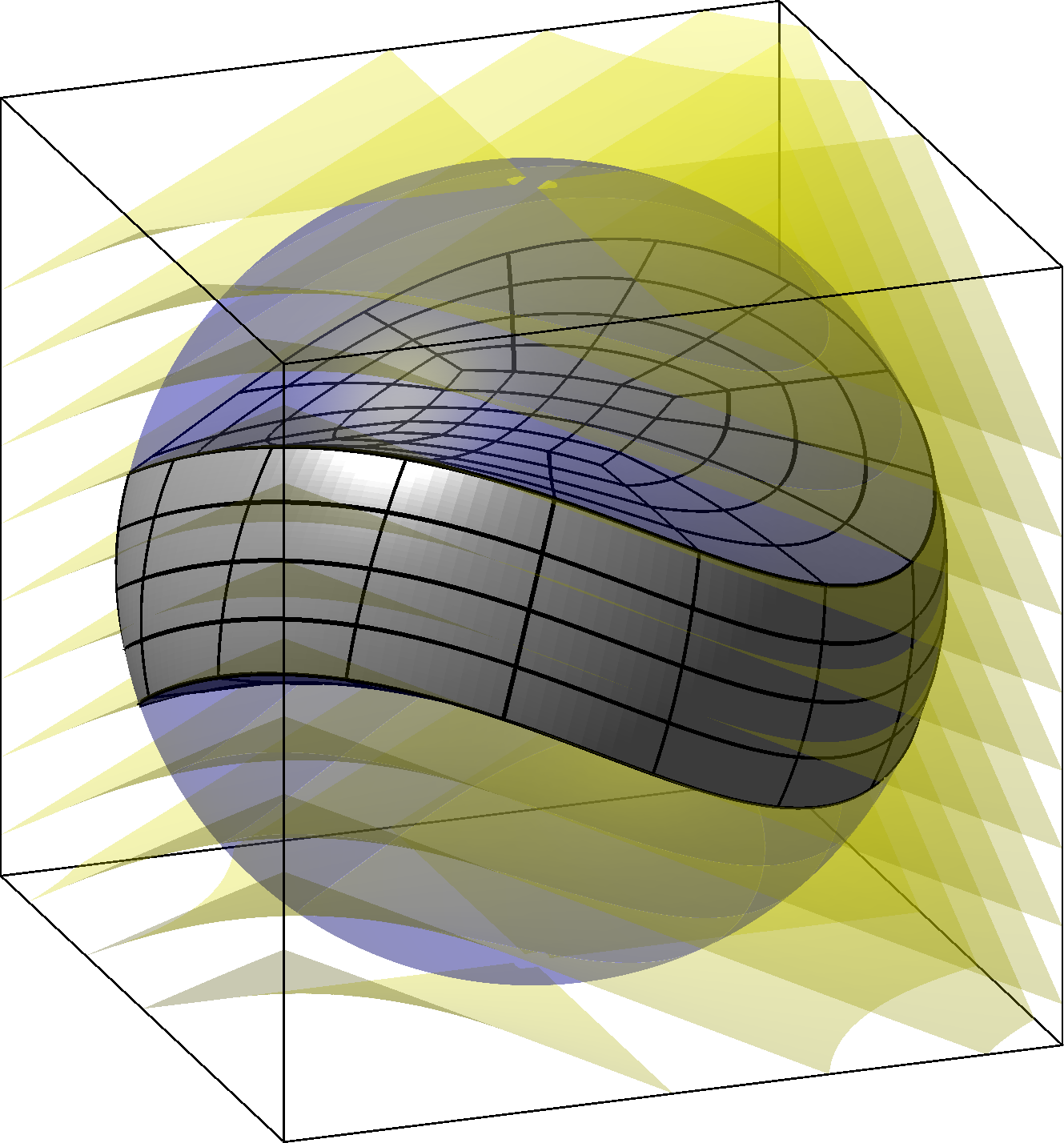}}\qquad\subfigure[discretized bulk domain and level sets]{\includegraphics[width=0.45\textwidth]{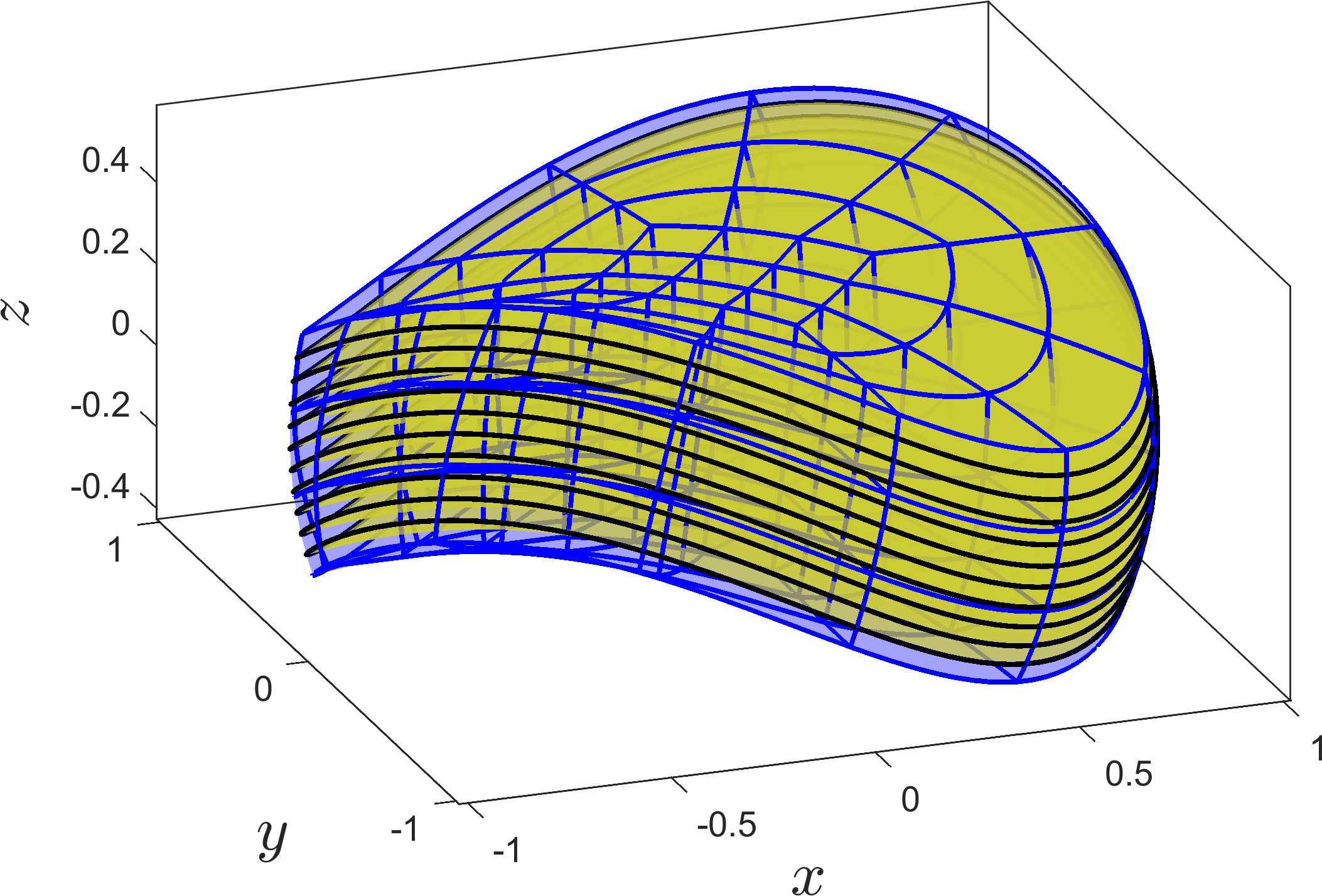}}
	
	\caption{\label{fig:TC11setup} Setup for the numerical example \ref{subsec:iTCns}: (a) The level-set function $\phi$ is shown in yellow, the whole sphere used to define the bulk domain $\Omega$ in light blue, and the considered interval-based bulk domain $\Omega$ is depicted with some discretization. (b) The discretized bulk domain $\Omega$ and some (arbitrarily selected) shells described by level sets $\Gamma^c$. Note that the shown meshes are just exemplarily chosen.}
\end{figure}

The bulk domain is prescribed as a subset of a sphere with $\phi^{\mathrm{min}}=-\nicefrac{1}{5}$ and $\phi ^{\mathrm{max}}=\nicefrac{2}{5}$, see Fig.~\ref{fig:TC11setup}. Therefore, the level-set function $\psi(\vek{x}) = \lVert \vek{x} \rVert - 1$ describing a sphere is introduced and then the bulk domain is defined as
\begin{equation}
	\Omega = \left\{\vek{x} \in \mathbb{R}^3\,:\, \psi(\vek{x}) \leq 0 \,\,\mathrm{and}\,\, -\nicefrac{1}{5} < \phi(\vek{x}) < \nicefrac{2}{5}\right\}.
\end{equation}

The material and geometric parameters for the shells are as follows: Young's modulus $E = 2.1 \cdot 10^7$, Poisson's ratio $\nu = 0.3$, and the shell thickness is $t = 0.1$. The shell surfaces are subjected to a loading described by body forces $\vek{f} = \left[0,0,-100\right]^{\mathrm{T}}$ and moments $\vek{c} = \vek{0}$. Navier support is prescribed along the whole boundary $\partial \Gamma^c$ of the shell.

\begin{figure}
	\centering
	
	\includegraphics[width=0.6\textwidth]{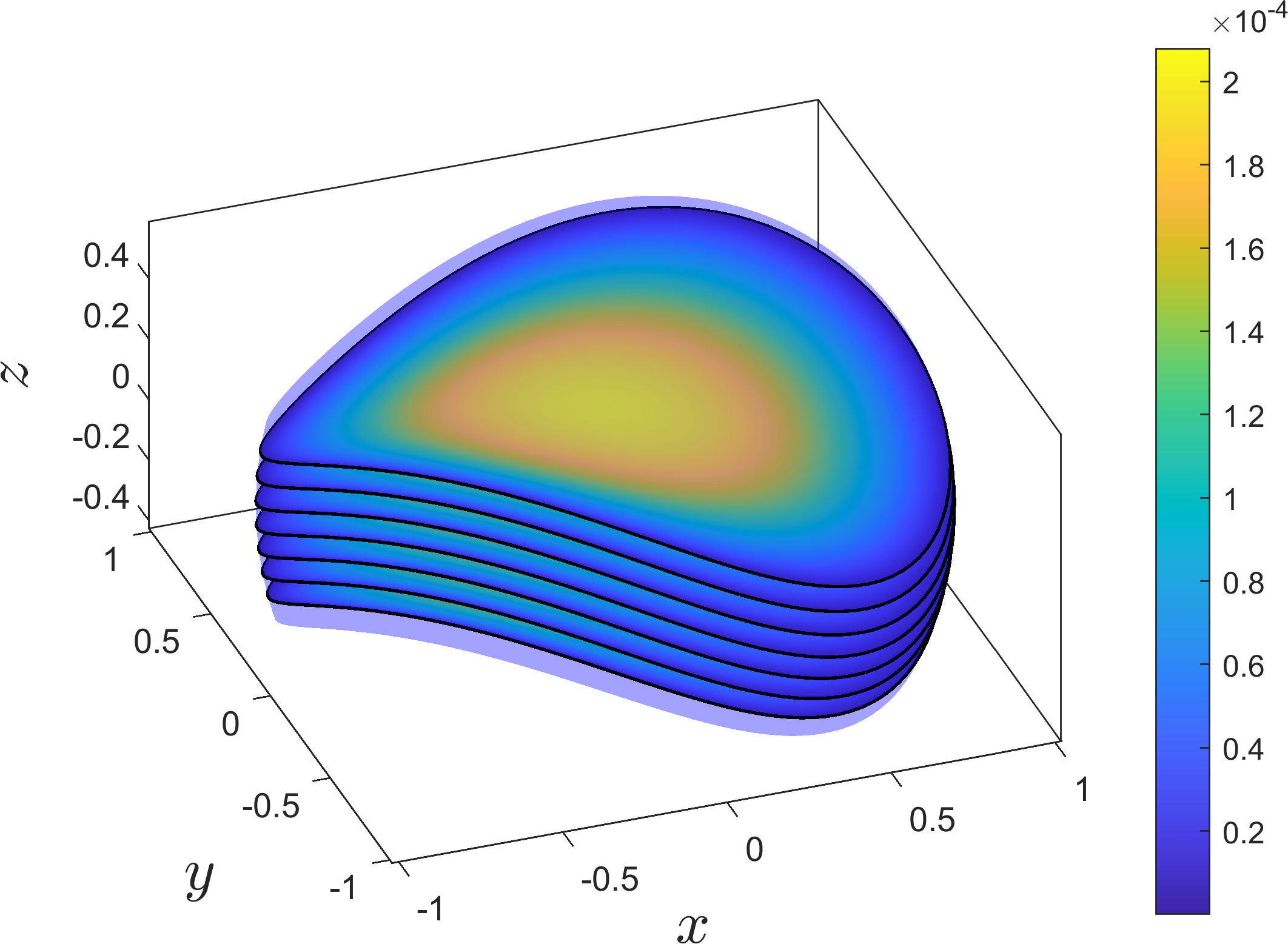}\hfill
	
	\caption{\label{fig:TC11res} Euclidean norm of the displacements $\lVert \vek{u} \rVert$. The light blue surface shows the envelop of the bulk domain including $\partial\Omega$.}
\end{figure}

\begin{figure}
	\centering
	
	\subfigure[convergence in $\varepsilon_{\mathrm{res,F}}$]{\includegraphics[width=0.32\textwidth]{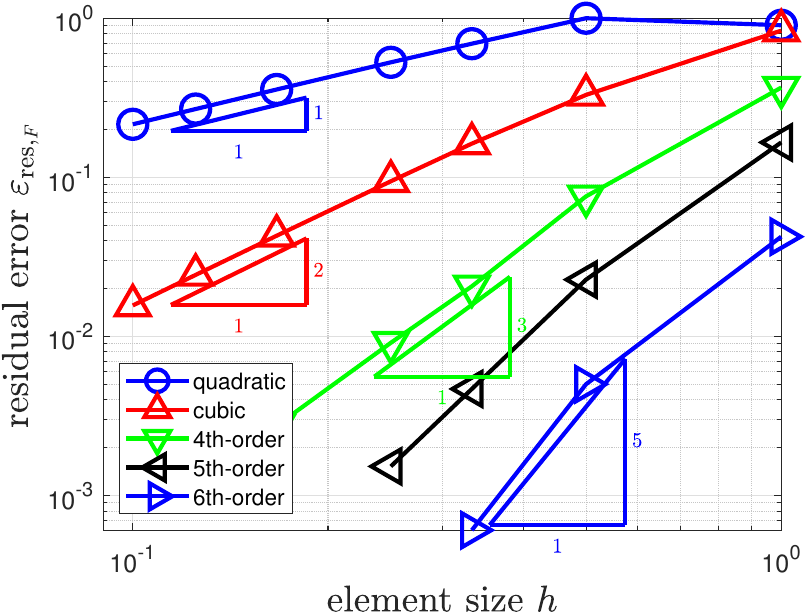}}\hfill\subfigure[convergence in $\varepsilon_{\mathrm{res,M}}$]{\includegraphics[width=0.32\textwidth]{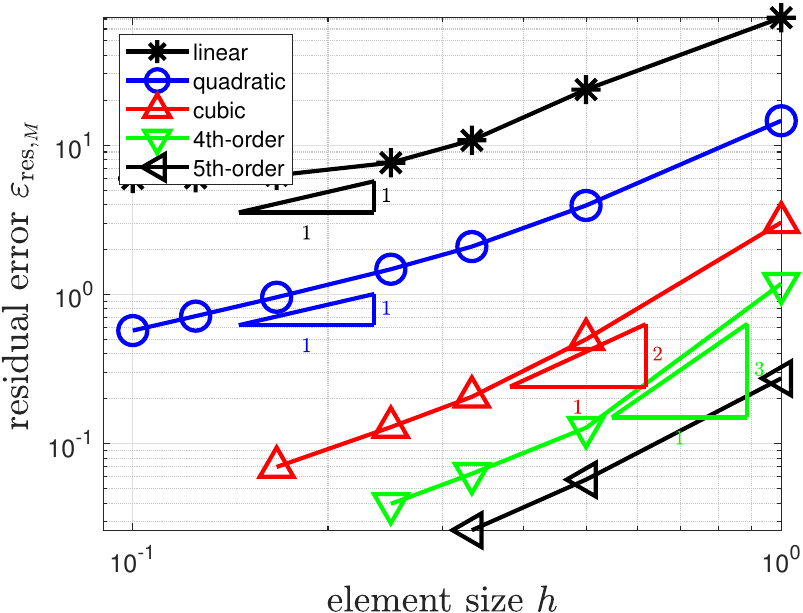}}\hfill\subfigure[convergence in $\varepsilon_{\mathfrak{e}}$]{\includegraphics[width=0.32\textwidth]{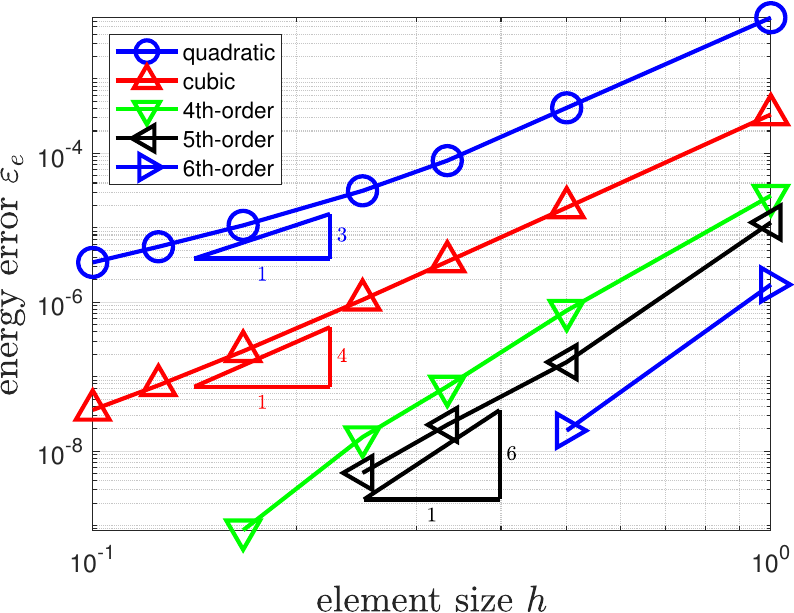}}
	
	\caption{\label{fig:TC11ConvStInterval} Convergence studies for the test case of trigonometric shells embedded in a spherical interval bulk domain. Convergence rates in (a) the residual error for the force equilibrium $\varepsilon_{\mathrm{res,F}}$, (b) the moment equilibrium $\varepsilon_{\mathrm{res,M}}$, and (d) convergence rates in the stored energy error $\varepsilon_{\mathfrak{e}}$.}
\end{figure}

The Euclidean norm of the displacements $\lVert \vek{u} \rVert$ is plotted over some arbitrarily selected shell geometries in Fig.~\ref{fig:TC11res}. Convergence studies for the error measures introduced in Section \ref{subsec:errorTypes} are shown in Fig.~\ref{fig:TC11ConvStInterval}. The convergence study in the residual error for the force equilibrium $\varepsilon_{\mathrm{res,F}}$ leads to optimal convergence rates $\mathcal{O}(p-1)$, see Fig.~\ref{fig:TC11ConvStInterval}(a), as described in Section \ref{subsec:errorTypes}. In the residual error $\varepsilon_{\mathrm{res,M}}$ for the moment equilibrium, the convergence rates are optimal for $p = \{2,3\}$, while for $p = \{4,5\}$ the curve flattens out for smaller element sizes $h$, see Fig.~\ref{fig:TC11ConvStInterval}(b), which is slightly sub-optimal. For the coupled system of PDEs, the convergence in the displacements of the middle surface of the shell are optimal, hence, the two convergence studies together are seen as a good confirmation for the proposed model and numerical method. Additionally, a convergence study in the stored energy error, Eq.~(\ref{eq:EnergyError}), also gives optimal convergence rates of at least $\mathcal{O}(p+1)$. The benchmark value for the convergence study in the stored energy norm is $\mathfrak{e}\left(\vek u\right)=9.917787434703 \cdot 10^{-3}$ and has been obtained by an overkill approximation using a very fine higher-order mesh. Similar investigations have been made in \cite{Fries_2023a} in the context of membranes. 

\subsection{Generalized test case with clamped support}\label{subsec:iTCcs}
\begin{figure}
	\centering
	
	\subfigure[level sets and bulk domain]{\includegraphics[width=0.3\textwidth]{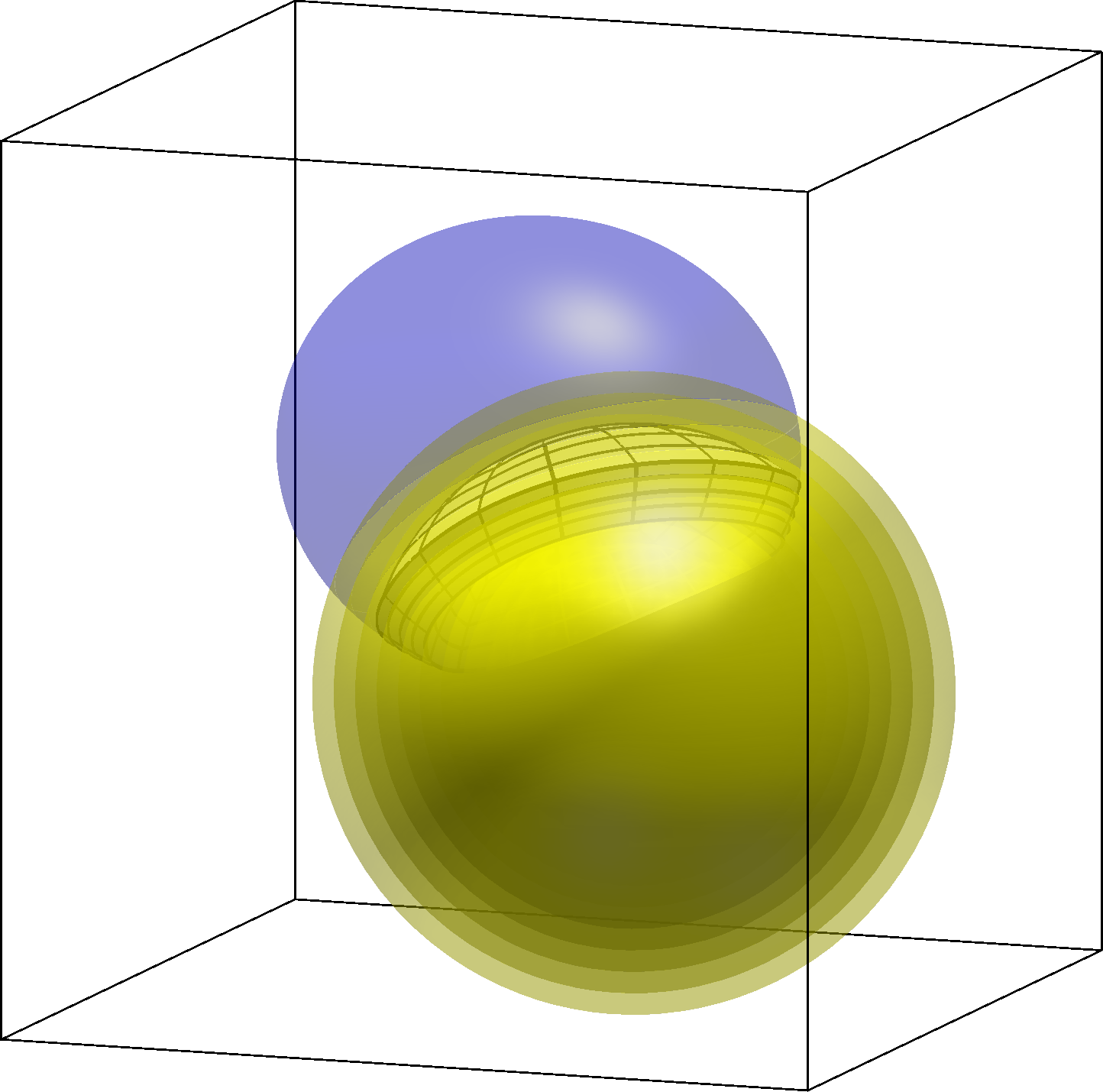}}\qquad\subfigure[discretized bulk domain and level sets]{\includegraphics[width=0.45\textwidth]{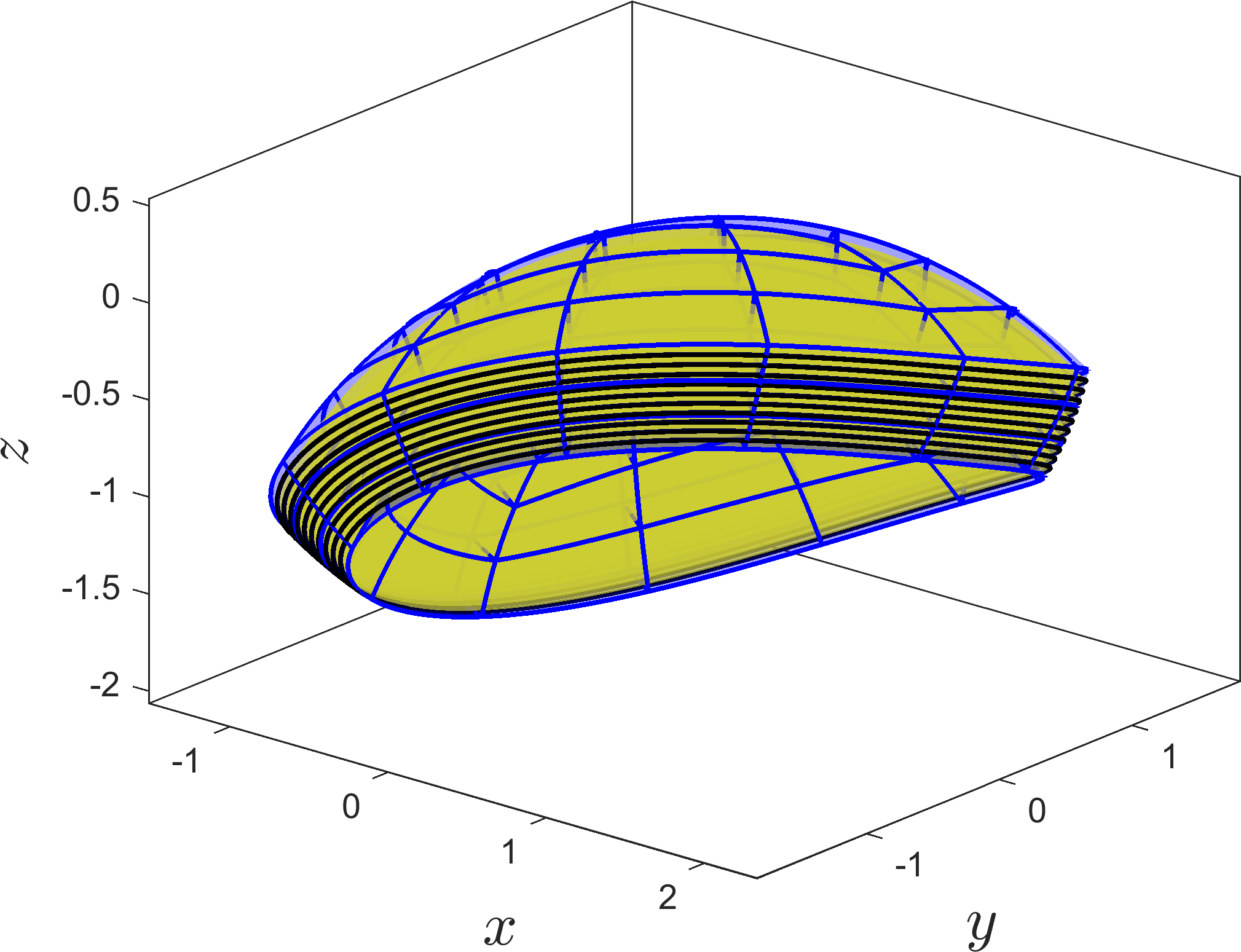}}
	
	\caption{\label{fig:TC23setup} Setup for the numerical example \ref{subsec:iTCns}: (a) The spherical level-set function $\phi$ is shown in yellow, the whole ellipsoid used to define the bulk domain $\Omega$ in light blue, and the considered interval-based bulk domain $\Omega$ with an example discretization. (b) The discretized bulk domain $\Omega$ and some (arbitrarily selected) shells described by level sets $\Gamma^c$.}
\end{figure}

In this example, spherical shells defined by the level-set function

\begin{equation}
	\phi(\vek{x}) = \lVert \vek{x} - \hat{\vek{x}}_\mathrm{C} \rVert - 2.
\end{equation}

with the center point $\hat{\vek{x}}_\mathrm{C} = \left[1, -0.5, -2\right]^{\mathrm{T}}$ are considered. For the bulk domain, first the level-set function
\[
\psi\left(\vek X\right)=\left(\frac{X-X_{E}}{R_{X}}\right)^{2}+\left(\frac{Y-Y_{E}}{R_{Y}}\right)^{2}+\left(\frac{Z-Z_{E}}{R_{Z}}\right)^{2}-2,
\]
with $\vek X_{\!E}=\left[-0.2,0.2,0.1\right]^{\mathrm{T}}$ and $R_{X}=1.25$,
$R_{Y}=1.00$, $R_{Z}=1.10$ is introduced as shown in blue in Fig.~\ref{fig:TC23setup}(a). Based on this, the bulk domain $\Omega$ is prescribed as
\begin{equation}
	\Omega = \left\{\vek{x} \in \mathbb{R}^3\,:\, \psi(\vek{x}) \leq 0 \,\,\mathrm{and}\,\, 0 < \phi(\vek{x}) < \nicefrac{1}{2}\right\}.
\end{equation}
The two level-set functions and the discretized bulk domain are shown in Fig.~\ref{fig:TC23setup}(a). In Fig.~\ref{fig:TC23setup}(b), this discretized bulk domain, with some embedded shells, is depicted in more detail and from a different perspective. Note that the shown number of elements in the mesh has been exemplarily chosen. The element number and order of the meshes have been systematically varied in the convergence studies. The material and geometric parameters for the shells are as follows: Young's modulus is $E = 10\,000$, Poisson's ratio is $\nu = 0.3$, and the shell thickness is $t = 0.05$. The shell surfaces are subjected to a loading described by body forces $\vek{f} = \left[0,0,-5\right]^{\mathrm{T}}$ and all boundaries of the shells are clamped.

\begin{figure}
	\centering

	\subfigure[]{\includegraphics[width=0.40\textwidth]{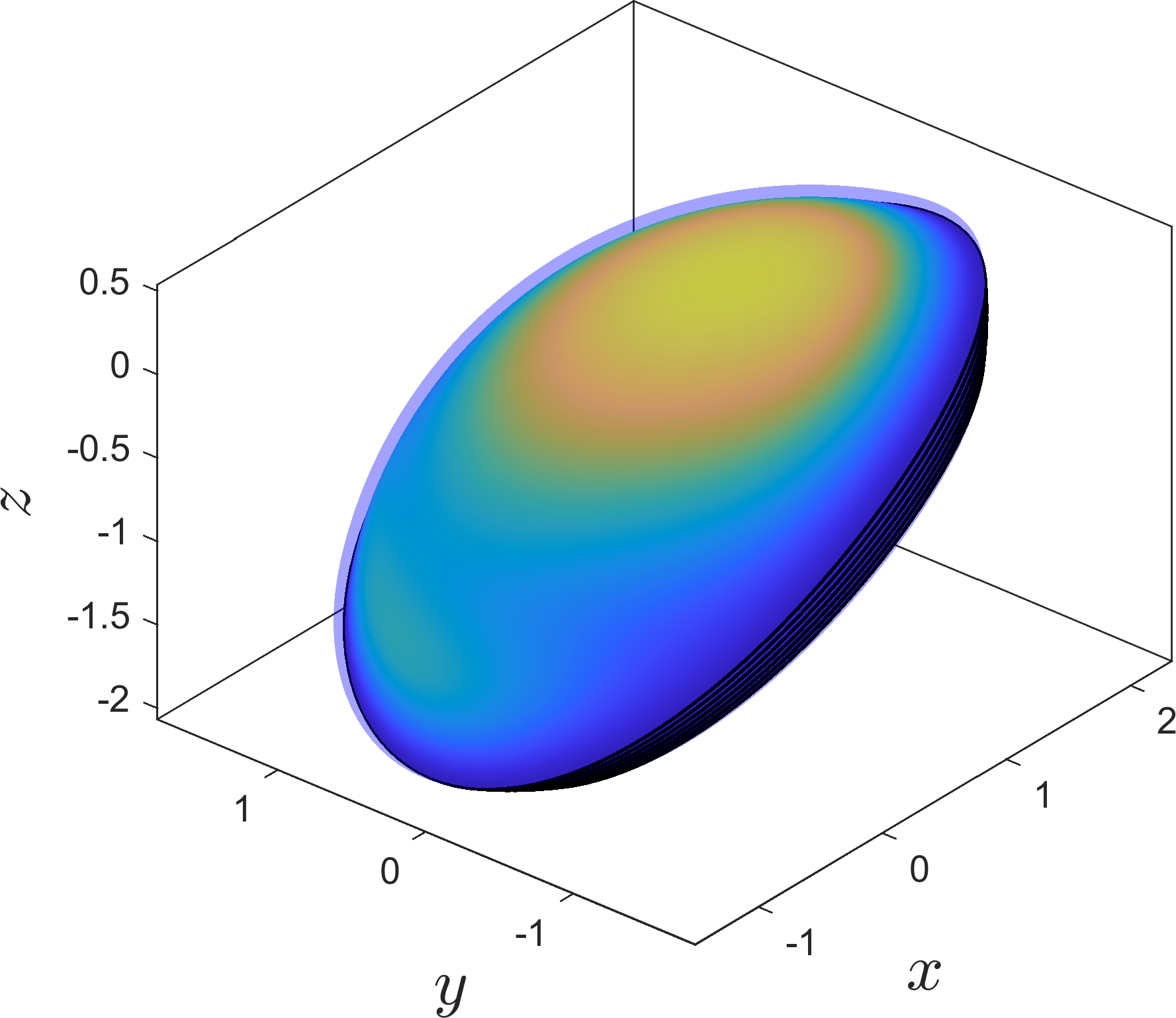}}\hspace{0.8cm}
	\subfigure[]{\includegraphics[width=0.48\textwidth]{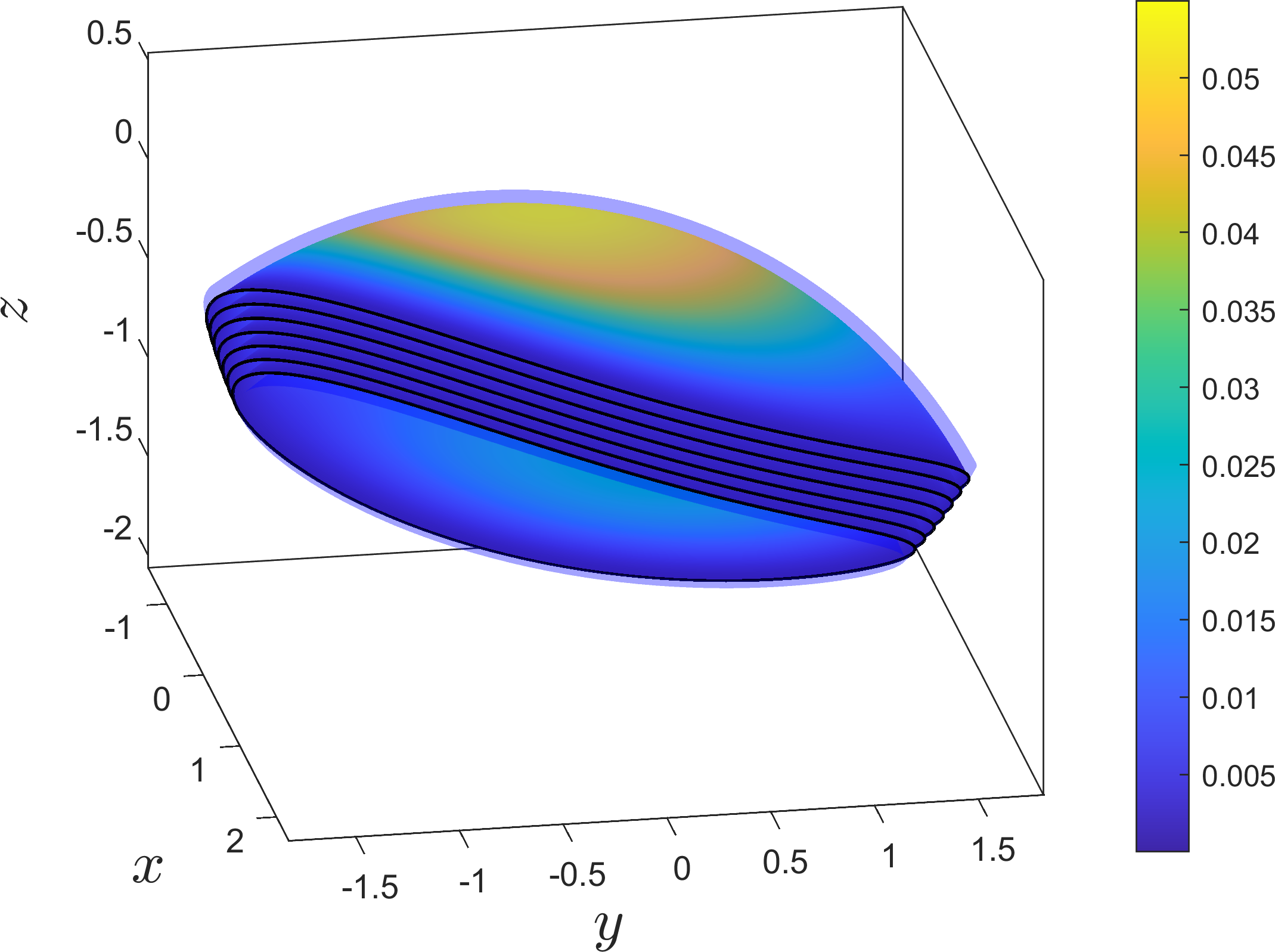}}
	
	\caption{\label{fig:TC23res} Euclidean norm of the displacements $\lVert \vek{u} \rVert$ plotted on some selected level sets $\Gamma^c$ depicted from two different perspectives. The light blue colour indicates the outer contour of $\Omega$.}
\end{figure}

\begin{figure}
	\centering
	
	\subfigure[convergence in $\varepsilon_{\mathrm{res,F}}$]{\includegraphics[width=0.32\textwidth]{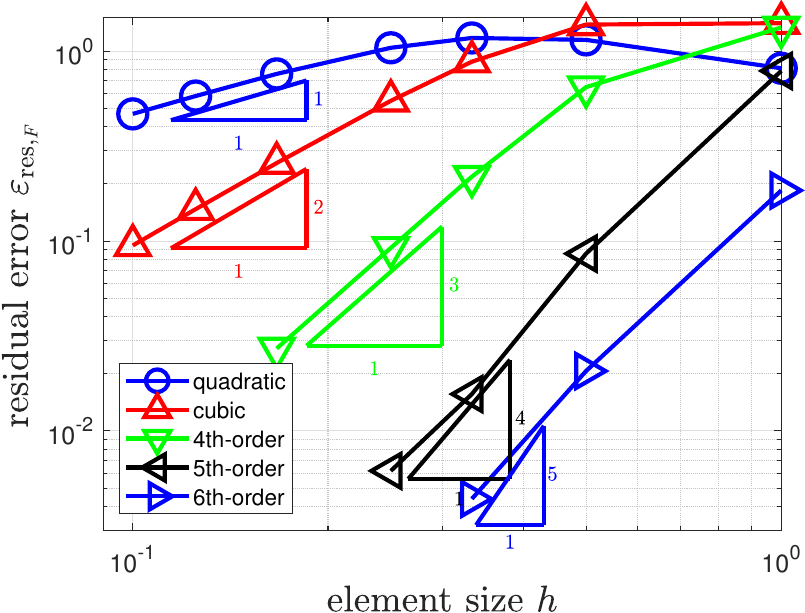}}\hfill\subfigure[convergence in $\varepsilon_{\mathrm{res,M}}$]{\includegraphics[width=0.32\textwidth]{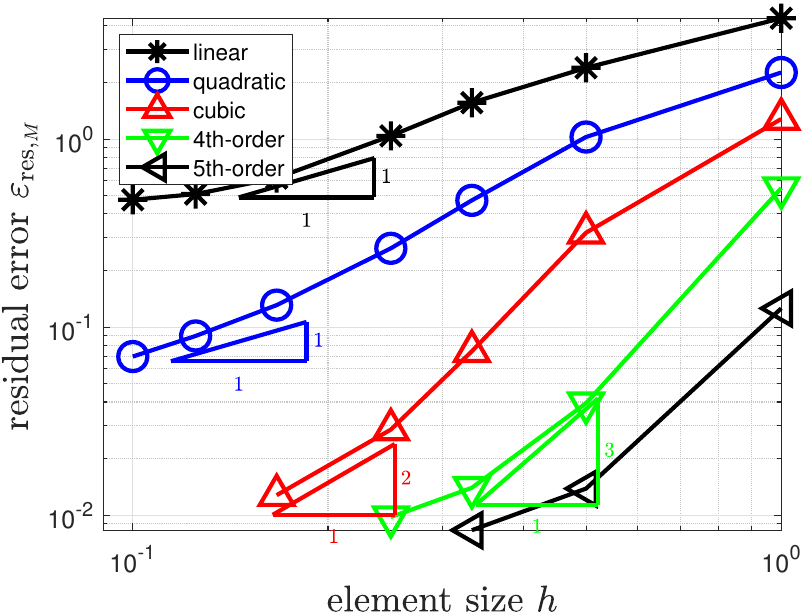}}\hfill\subfigure[convergence in $\varepsilon_{\mathfrak{e}}$]{\includegraphics[width=0.32\textwidth]{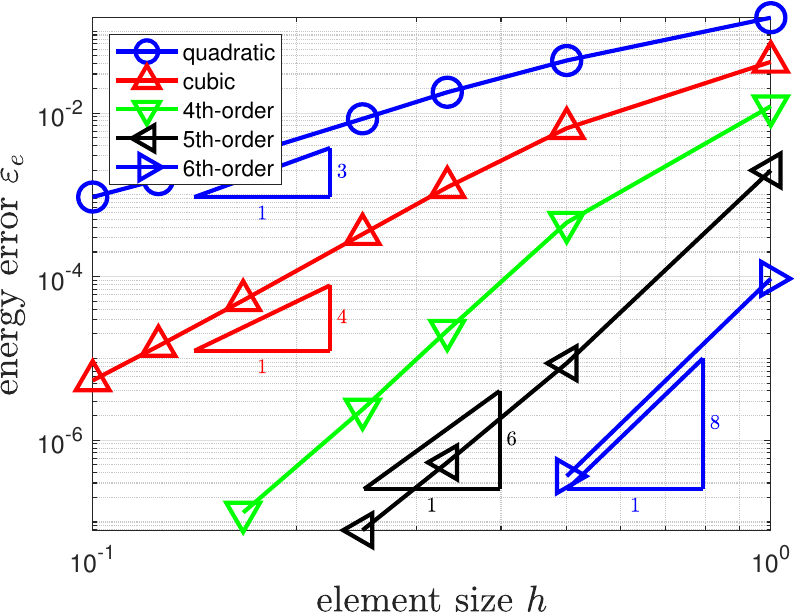}}
	
	\caption{\label{fig:TC23ConvStInterval} Convergence studies for the test case of clamped spherical shells embedded in an ellipsoidal bulk domain: (a) convergence rates in the strong form error for the force equilibrium $\varepsilon_{\mathrm{res,F}}$, and (b) the moment equilibrium $\varepsilon_{\mathrm{res,M}}$; (d) convergence rates in the stored energy error $\varepsilon_{\mathfrak{e}}$.}
\end{figure}

The Euclidean norm of the displacement field is shown in Fig.~\ref{fig:TC23res}. The benchmark value for the convergence study in the stored energy norm is $\mathfrak{e}\left(\vek u\right)=0.3479397986834$ and has been obtained by an overkill approximation. Fig.~\ref{fig:TC23ConvStInterval} shows higher-order convergence results in the residual errors and the stored energy error for this example. The results are similar to the example shown in Section \ref{subsec:iTCns}. In the residual error of the force equilibrium, optimal convergence rates are obtained, while for the moment equilibrium, the curves flatten out for higher-order elements. However, the performance of both types of residual errors for the coupled system of PDEs confirms that the mechanical model and the applied FEM are correctly formulated and implemented, respectively. The convergence study in the stored energy error shows the estimated convergence behaviour for all element orders $p$.

\subsection{Cupolas on a curved surface with clamped supports \label{subsec:DVS-clamped}}

\begin{figure}
	\centering
	
	\includegraphics[width=0.6\textwidth]{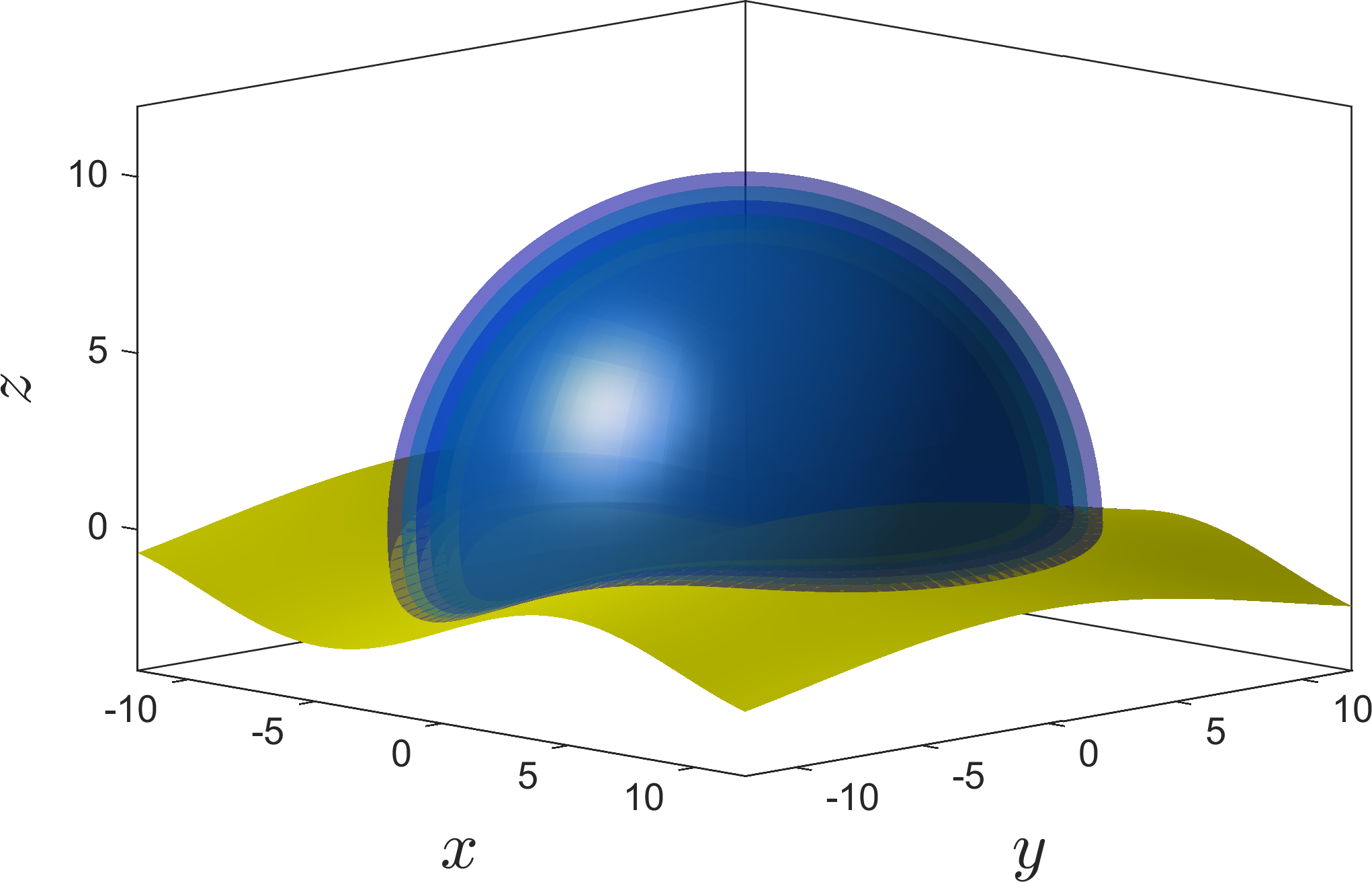}\hfill
	
	\caption{\label{fig:CupTC} Half-spherical bulk domain with some embedded cupolas placed on the surface implied by the level-set function $\psi$ depicted in yellow.}
\end{figure}

\begin{figure}
	\centering
	
	\subfigure[residual error in force equilibrium]{\includegraphics[width=0.5\textwidth]{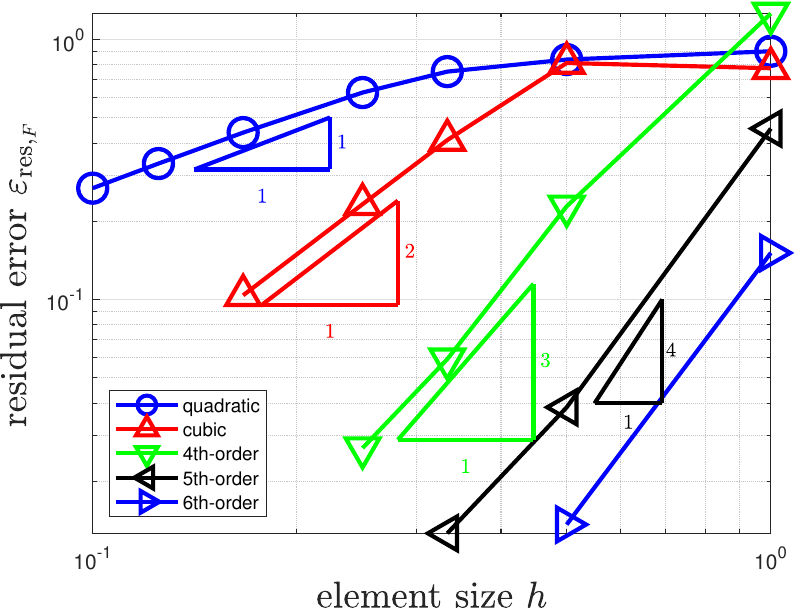}}\hfill\subfigure[residual error in moment equilibrium]{\includegraphics[width=0.5\textwidth]{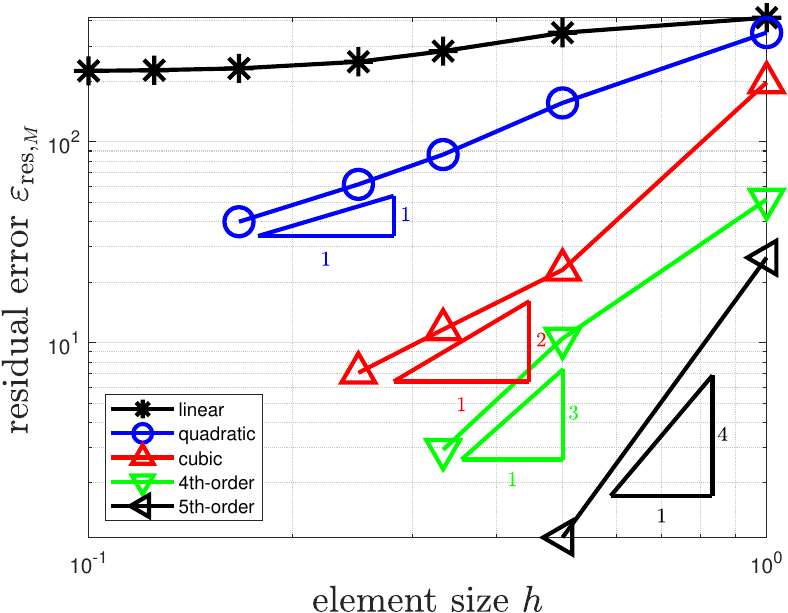}}	
	\caption{\label{fig:ResConvStud-ClampSup} Higer-order convergence study in the residual error for the half-spherical cupola on a curved surface with completely clamped support along the whole boundary: Convergence rates in the strong form error for the (a) force equilibrium and the (b) moment equilibrium.}
\end{figure}

\begin{figure}
	\centering
	
	\includegraphics[width=0.6\textwidth]{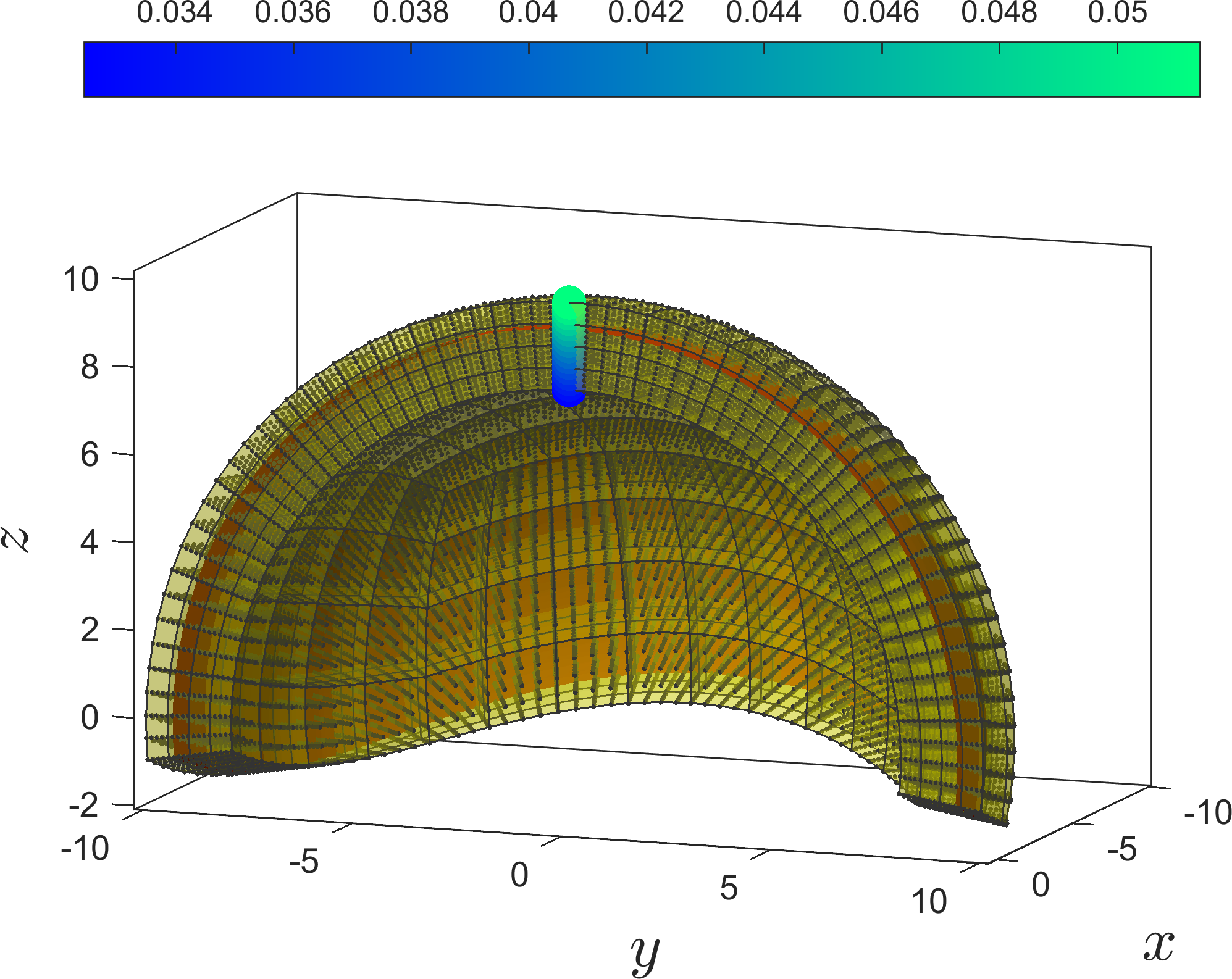}\hfill
	
	\caption{\label{fig:DesignValueSearch} Some higher-order mesh to find the target value cut along $x = 0$. The colours at the apex nodes indicate the considered normalized deformation $\lVert \vek{u} \rVert$. The red surface is the obtained midsurface of the shell which fulfils the target value $u_{\mathrm{DV}}$.}
\end{figure}

\begin{figure}
	\centering 
	\includegraphics[width=0.6\textwidth]{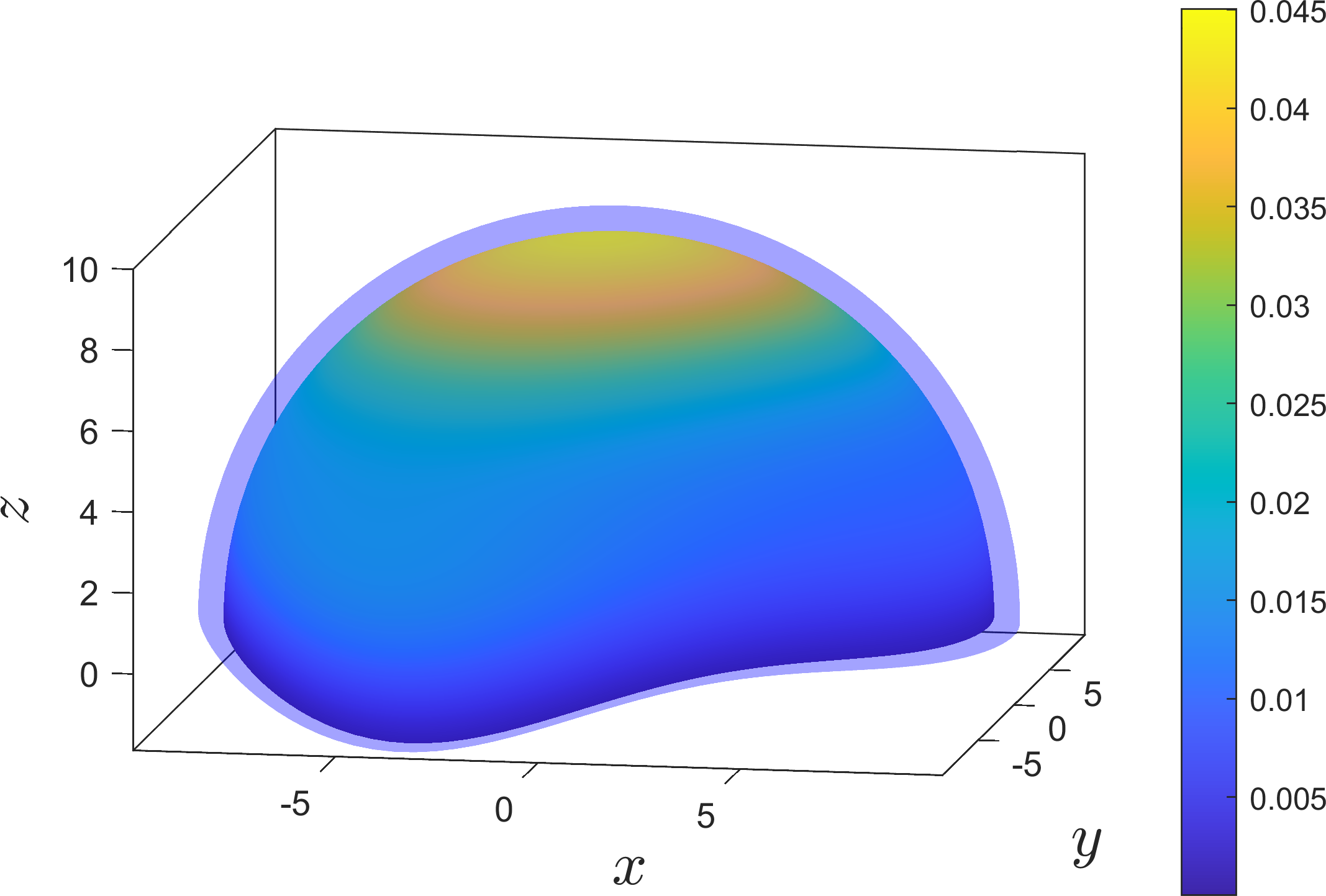}
	\caption{\label{fig:SolSrfClamped} The Euclidean norm of the displacements $\lVert \vek{u} \rVert$ plotted over the isosurface with radius $r_{\mathrm{t}} = 9.3770\,\unit{m}$ within the bulk domain. The light blue surface indicates the contour of the bulk domain.}
\end{figure}

In this example, a half-spherical cupola is placed on a curved surface implied by the level set function $\psi = z - (\sin(\frac{x}{3}) + \cos(\frac{y}{6})) + 1.0$ as shown in Fig.~\ref{fig:CupTC}. The whole boundary of the cupola on the surface is clamped. The half-spherical bulk domain has an outer radius of $R_{\mathrm{o}} = 10.0\,\unit{m}$ and an inner radius of $R_{\mathrm{i}} = 8.0\,\unit{m}$. Young's modulus is defined as $E = 30\,000 \,\unit{MPa}$, Poisson's ratio as $\nu = 0.3$, the load in vertical direction is $f_z =- 10\,\unit{MN/m^2}$ (acting downwards), and the assumed shell thickness associated to every level set is $t=1.10\,\unit{m}$, resulting in a moderately thick shell. Hence, this test case serves more as a real-world example. Analogously to the more academic test cases from above, convergence studies in the residual error for the force and moment equilibrium have been performed, respectively. Fig.~\ref{fig:ResConvStud-ClampSup} shows these convergences studies which confirm optimal convergence rates, in this case for the residual errors with respect to the force and moment equilibrium. Optimal convergence rates as estimated has been achieved. The convergence rate of the $6^{\mathrm{th}}$-order mesh in the force equilibrium may still be in the pre-asymptotic ranch and, therefore, not achieve $5^{\mathrm{th}}$-order convergence as expected. However, the convergence rates of the moment equilibrium do not face the problems of flattening out as the test cases in Sections \ref{subsec:iTCns} and \ref{subsec:iTCcs} do.\\
\\
It is now shown how the proposed modelling may be used in the structural design of some shell. Let there be a target design value, e.g., for some maximum displacement. In the present test case, this design value (DV) is a prescribed value for the Euclidean norm $ u_{\mathrm{DV}} =  \lVert \vek{u} \rVert = 0.045$ of the displacements along the $z$-axis ($x=y=0$). The aim is to find the middle surface geometry (i.e., the isosurface) with the smallest radius such that the target value is not exceeded. Therefore, the values for $\lVert \vek{u} \rVert$ at the nodes $(0,0,z)$ within the bulk domain mesh are evaluated and then the radius yielding the target design value is extracted by interpolation. Fig.~\ref{fig:DesignValueSearch} shows some mesh which is cut along the $x$-axis to visualize the deformation of the nodes at $(0,0,z)$ and the corresponding normalized deformation $ \lVert \vek{u} \rVert = \sqrt{u^2 + v^2 + w^2}$. The obtained minimal radius for the middle surface of the shell for which $u_{\mathrm{DV}}$ occurs is $r_{\mathrm{t}} = 9.3770\,\unit{m}$ and shown as the red surface in Fig.~\ref{fig:DesignValueSearch}. Fig.~\ref{fig:SolSrfClamped} shows $ \lVert \vek{u} \rVert $ on this isosurface defined by $r_{\mathrm{t}}$ within the bulk domain.

\section{Conclusions and outlook}\label{sec:Concl}

A mechanical model for linear Reissner--Mindlin shells is formulated to simultaneously consider \emph{all} shells described by level sets over a bulk domain. For the formulation of this model and the corresponding numerical method, several geometric quantities (vector fields, curvature) and differential surface operators are required. These are based on the TDC which leads to a coordinate-free formulation of the governing equations for the mechanical quantities in the shells and readily applies to implicitly defined geometries, e.g., the level sets used in this work. Possible applications for this approach are the search for extreme values of mechanical fields (e.g., stresses) for a whole family of shell geometries and the modelling of anisotropic materials which includes shells as reinforcing sub-structures.\\
\\
For the numerical analysis, the Bulk Trace FEM, introduced by the authors in \cite{Fries_2023a}, is applied in the present context. The employed meshes are conforming to the boundary of the bulk domain and, therefore, also to the boundaries of the embedded level sets, i.e., the shells. However, the discretization does, in fact, \emph{not} conform to the middle surfaces of the shells, i.e., the shape of the level sets.  In this sense, properties of classical conforming FEMs and non-conforming fictitious domain methods (e.g., the Trace FEM) are combined in this method; hence, it may be labelled Bulk Trace FEM. A mixed ansatz for the shape functions in the FEM is used for the first time in the Bulk Trace FEM for structural mechanics. Simple support conditions, e.g., clamped edges or Navier supports may be enforced strongly by prescribing nodal values. For more advanced situations of essential (Dirichlet) boundary conditions, a weak formulation may be useful. Herein, the non-symmetric Nitsche's method was used to prescribe symmetry boundary conditions.\\
\\
Numerical examples based on classical benchmark test cases for single shell geometries and more general ones, which enable higher-order convergence studies, are shown. Based on these numerical examples, it was confirmed that the proposed mechanical model and the Bulk Trace FEM are valid for the simultaneous solution of shells embedded in a higher-dimensional bulk domain. We believe that this approach is a powerful extension of existing structural shell mechanics for single shell geometries.\\
\\
Further research shall focus on the application of this approach, e.g., in more advanced design value searches, integration of this method in optimization strategies used in structural design processes, and in new, anisotropic models with embedded sub-structures. In addition to the ropes and membranes considered in \cite{Fries_2023a} and the linear Reissner--Mindlin shells presented herein, the formulation of curved beams, e.g., curved Timoshenko beams, shall be developed, e.g., to the influence of embedded fibres with bending resistance in bulk materials.

\bibliographystyle{schanz}
\bibliography{C:/Users/micha/Documents/IFB-Documents-MWK/SciWriting/Sci-Lit-Management/MWKpubRefs.bib}
 
\end{document}